\documentclass[10pt,a4paper,twoside]{article}
\usepackage{RR}
\usepackage[latin1]{inputenc}
\usepackage{amsfonts,amsmath,amssymb,amsthm,amstext,latexsym,paralist}
\usepackage{url}
\usepackage{xspace}

\usepackage{a4wide}
\usepackage{algorithm2e}

\usepackage{figlatex}

\newtheorem{lemma}{Lemma}
\newtheorem{theorem}{Theorem}
\newtheorem{proposition}{Proposition}
\newtheorem{corollary}{Corollary}

\AtBeginDocument{}

\newcommand{\server}{\textsf{server}}
\newcommand{\Servers}{\textsf{Servers}}
\newcommand{\Ancests}{\textsf{Ancestors}}

\newcommand{\sto}{\textsf{sc}}  
\newcommand{\dist}{\textsf{dist}}
\newcommand{\comm}{\textsf{comm}}
\newcommand{\comp}{\textsf{comp}}
\newcommand{\II}{\mathcal{I}} 
\newcommand{\TT}{\mathcal{T}} 
\newcommand{\NN}{\mathcal{N}} 
\newcommand{\NNN}{\mathcal{N}'} 
\newcommand{\CC}{\mathcal{C}} 
\newcommand{\LL}{\mathcal{L}} 
\newcommand{\qos}{\textsf{q}}
\newcommand{\W}{\textsf{W}} 

\newcommand{\rr}{r}

\newcommand{\CL}{\textsf{BW}} 
\newcommand{\BW}{\textsf{BW}}
\newcommand{\parent}{\textsf{parent}}
\newcommand{\child}{\textsf{children}}
\newcommand{\clients}{\textsf{clients}}
\newcommand{\inreq}{\textsf{inreq}}
\newcommand{\treated}{\textsf{treated}}
\newcommand{\replica}{\textsf{replica}}
\newcommand{\flow}{\textsf{flow}} 
\newcommand{\suc}{\textsf{succ}}

\newcommand{\subtree}{\textsf{subtree}}

\newcommand{\pat}[2]{\textsf{path}[{#1} \to {#2}]}
\newcommand{\REP}{\textsc{Replica Placement}\xspace}
\newcommand{\COR}{\textsc{Replica Cost}\xspace}
\newcommand{\RCO}{\textsc{Replica Counting}\xspace}
\newcommand{\CLOSEST}{\textit{Closest}\xspace}
\newcommand{\UPWARDS}{\textit{Upwards}\xspace}
\newcommand{\MULTIPLE}{\textit{Multiple}\xspace}

\newcommand{\ctdall}{CTDA\xspace}

\newcommand{\ctdlf}{CTDLF\xspace}

\newcommand{\cbu}{CBU\xspace}
\newcommand{\ubu}{UBCF\xspace}
\newcommand{\utd}{UTD\xspace}
\newcommand{\mtd}{MTD\xspace}
\newcommand{\mbu}{MBU\xspace}
\newcommand{\mg}{MG\xspace}

\newcommand{\utdFirstPass}{UTDFirstPass\xspace}
\newcommand{\utdSecondPass}{UTDSecondPass\xspace}

\newcommand{\mbuFirstPass}{MBUFirstPass\xspace}
\newcommand{\mbuSecondPass}{MBUSecondPass\xspace}

\RRetitle{Strategies for Replica Placement in Tree Networks}
\RRtitle{Stratégies de placement de répliques sur des arbres}

\RRdate{November 2006}

\RRauthor{Anne Benoit \and Veronika Rehn \and Yves Robert}
\titlehead{Strategies for Replica Placement in Tree Networks}
\authorhead{A. Benoit, V. Rehn, Y. Robert}

\RRabstract{In this paper, we discuss and compare several policies to place
replicas in tree networks, subject to server capacity and QoS
constraints. The client requests are known beforehand, while the
number and location of the servers are to be determined. The
standard approach in the literature is to enforce that all requests
of a client be served by the closest server in the tree. We
introduce and study two new policies. In the first policy, all
requests from a given client are still processed by the same server,
but this server can be located anywhere in the path from the client
to the root. In the second policy, the requests of a given client
can be processed by multiple servers.

One major contribution of this paper is to assess the impact of
these new policies on the total replication cost. Another important
goal is to assess the impact of server heterogeneity, both from a
theoretical and a practical perspective. In this paper, we establish
several new complexity results, and provide several efficient
polynomial heuristics for NP-complete instances of the problem.
These heuristics are compared to an absolute lower bound provided by
the formulation of the problem in terms of the solution of an
integer linear program.}
\RRkeyword{Replica placement, tree networks, access policy, scheduling, complexity results,
 heuristics, heterogeneous clusters.}

\RRresume{Dans ce rapport nous présentons et comparons plusieurs politiques
de placement de répliques sur des arbres, prenant en compte à la fois des contraintes
liées à la capacité de traitement
de chaque serveur et des contraintes de type QoS (qualité de service). Les
requêtes des clients sont
connues avant exécution, alors que le nombre et l'emplacement des répliques (serveurs)
sont à déterminer par l'algorithme de placement.
L'approche classique impose que toutes les requêtes d'un client donné soient
traitées par un seul serveur, à savoir le plus proche du client dans l'arbre.
Nous introduisons deux nouvelles politiques de placement. Dans la première,
chaque client a toujours un serveur unique, mais ce dernier peut être
situé n'importe où sur le chemin qui mène du client à la racine dans l'arbre.
Avec la deuxième politique, les requêtes d'un même client peuvent être traitées par
plusieurs serveurs sur ce même chemin.

Nous montrons que ces deux nouvelles
politiques de placement sont à même de réduire fortement
le coût total de la réplication.
Un autre objectif de ce travail est l'analyse
de l'impact de l'hétérogénéité de la plate-forme,
à la fois d'un point de vue théorique et pratique.
Sur le plan théorique, nous établissons plusieurs résultats de complexité,
dans les cadres homogène et hétérogène, pour l'approche classique
et les nouvelles politiques.
Sur le plan pratique, nous
concevons des heuristiques polynomiales pour les instances combinatoires du problème.
Nous comparons les performances de ces heuristiques en les
rapportant à une borne inférieure absolue
sur le coût total de la réplication;
cette borne est
obtenue par relaxation d'un programme linéaire en nombre entiers qui
caractérise la solution optimale du problème.}
\RRmotcle{Placement de répliques, réseaux en arbre, ordonnancement, complexité,
heuristiques, grappes de calcul hétérogènes.}

\RRtheme{\THNum}
\URRhoneAlpes
\RRprojet{GRAAL}

\begin{document}

\makeRR

\tableofcontents
\newpage




\section{Introduction}
\label{sec.intro}

In this paper, we consider the general problem of replica placement in tree
networks. Informally, there are clients issuing requests to be satisfied by servers.
The clients are known (both their position in the tree and their number of requests), while the
number and location of the servers are to be determined. A client is a leaf node
of the tree, and its requests can be served by one or several internal nodes.
Initially, there are no replica; when a node is equipped with a replica, it can process a number
of requests, up to its capacity limit. Nodes equipped with a
replica, also called servers, can only serve clients located in
their subtree (so that the root, if equipped with a replica, can serve any client); this restriction is
usually adopted to enforce the hierarchical nature of the target application platforms, where a node
has knowledge only of its parent and children in the tree.

The rule of the game is to assign replicas to nodes
so that some optimization function is minimized.
Typically, this optimization function is the total utilization cost
of the servers.
If all the nodes are identical, this reduces to minimizing
the number of replicas. If the nodes are heterogeneous, it is natural to assign a cost proportional to their capacity
(so that one replica on a node capable of handling $200$ requests is equivalent to two replicas on nodes
of capacity $100$ each).

The core of the paper is devoted to the study of the previous optimization problem, called
\REP in the following. Additional constraints
are introduced, such as guaranteeing some Quality of Service
(QoS): the requests must be served
in limited time, thereby prohibiting too remote or hard-to-reach
replica locations. Also, the flow
of requests through a link in the tree cannot exceed some
bandwidth-related capacity.
We focus on optimizing the total utilization cost (or replica number in the homogeneous case). There is a bunch
of possible extensions: dealing with several object types rather than one, including communication time
into the objective function, taking into account an update cost of the replicas, and so on. For the sake
of clarity we devote a special section (Section~\ref{sec.extensions}) to formulate these extensions,
and to describe which situations our results and algorithms can still apply to.

We point out that the distribution tree (clients and nodes) is fixed in our approach.
This key assumption is quite natural for a broad spectrum of applications, such as electronic, ISP, or VOD
service delivery. The root server has the original copy of the database
but cannot serve all clients directly,
so a distribution tree is deployed to provide a hierarchical and distributed access to replicas of the
original data. On the contrary, in other, more decentralized, applications
(e.g. allocating Web mirrors in distributed
networks), a two-step approach is used: first determine a ``good'' distribution
tree in an arbitrary interconnection graph, and then determine a ``good'' placement of replicas among the tree
nodes. Both steps are interdependent, and the problem is much more complex, due to the combinatorial
solution space (the number of candidate distribution trees may well be exponential).

Many authors deal with the \REP optimization problem, and we survey related work in Section~\ref{sec.related}.
The objective of this paper is twofold: (i) introducing two new access policies and comparing them with the
standard approach; (ii) assessing the impact of server heterogeneity on the problem.

In most, if not all, papers from the literature, all requests of a client are served by the closest replica,
i.e. the first replica found in the unique path from the client to the root in the distribution tree. This
\CLOSEST policy is
simple and natural, but may be unduly restrictive, leading to a waste of resources. We introduce and
study two different approaches: in the first one, we keep the restriction that all requests from a given client
are processed by the same replica, but we allow client requests to ``traverse'' servers so as to be processed
by other replicas located higher in the path (closer to the root). We call this approach the \UPWARDS
policy. The trade-of to explore is the following: the \CLOSEST policy assigns replicas at proximity of the
clients, but may need to allocate too many of them if some local subtree issues a great number of requests.
The \UPWARDS policy will ensure a better resource usage, load-balancing the process of requests on a larger
scale; the possible drawback is that requests will be served by remote servers, likely to take longer time
to process them. Taking QoS constraints into account would typically be more important for the \UPWARDS
policy.

In the second approach, we further relax access constraints and grant
the possibility for a client to be assigned
several replicas. With this \MULTIPLE policy, the processing
of a given client's requests will be split among
several servers located in the tree path from the client
to the root. Obviously, this policy is the most
flexible, and likely to achieve the best resource usage.
The only drawback is the (modest) additional complexity
induced by the fact that requests must now be tagged with the
replica server ID in addition to the client ID.
As already stated, one major objective of this paper
is to compare these three access policies,
\CLOSEST, \UPWARDS and \MULTIPLE.

The second major contribution of the paper is to assess the impact
of server heterogeneity, both from a theoretical and a practical
perspective. Recently, several variants of the \REP optimization
problem with the \CLOSEST policy have been shown to have polynomial
complexity. In this paper, we establish several new complexity
results. Those for the homogeneous case are surprising: for the
simplest instance without QoS nor bandwidth constraints, the
\MULTIPLE policy is polynomial (as \CLOSEST) while \UPWARDS is
NP-hard. The three policies turn out to be NP-complete for
heterogeneous nodes, which provides yet another example of the
additional difficulties induced by resource heterogeneity. On the
more practical side, we provide an optimal algorithm for the
\MULTIPLE problem with homogeneous nodes, and several heuristics for
all three policies in the heterogeneous case. We compare these
heuristics through simulations conducted for problem instances
without QoS nor bandwidth constraints. Another contribution is that
we are able to assess the absolute performance of the heuristics,
not just comparing one to the other, owing to a lower bound provided
by a new formulation of the \REP problem in terms of an integer
linear program: the relaxation of this program to the rational
numbers provides a lower bound to the solution cost (which is not
always feasible).

The rest of the paper is organized as follows.
Section~\ref{sec.model} is devoted to a detailed presentation of the
target optimization problems. In Section~\ref{sec.strategies} we
introduce the three access policies, and we give a few motivating
examples. Next in Section~\ref{sec.complexity} we proceed to the
complexity results for the simplest version of the \REP problem,
both in the homogeneous and heterogeneous cases.
Section~\ref{sec.lp} deals with the formulation for the \REP problem
in terms of an integer linear program. In
Section~\ref{sec.heuristics} we introduce several polynomial
heuristics to solve the \REP problem with the different access
policies. These heuristics are compared through simulations, whose
results are analyzed in Section~\ref{sec.experiments}.
Section~\ref{sec.extensions} discusses various extensions to the
\REP problem while Section~\ref{sec.related} is devoted to an
overview of related work. Finally, we state some concluding remarks
in Section~\ref{sec.conclusion}.

\section{Framework}
\label{sec.model}

This section is devoted to a precise statement of the \REP optimization problem.
We start with some definitions and notations. Next we outline the simplest instance
of the problem. Then we describe several types of
constraints that can be added to the formulation.

\subsection{Definitions and notations}

We consider a distribution tree $\TT$ whose nodes are partitioned into a set of clients $\CC$
and a set of nodes $\NN$. The set of tree edges is denoted as $\LL$.
The clients are leaf nodes of the tree, while $\NN$ is the set of
internal nodes.
It would be easy to
allow \emph{client-server} nodes which play both the rule of a
client and of an internal node (possibly a server),
by dividing such a node into two distinct nodes in the
tree, connected by an edge with zero communication cost.

A \emph{client} $i \in \CC$ is making requests to database objects. For the sake of clarity,
we restrict the presentation to a single object type, hence a single database. We deal with
several object types in Section~\ref{sec.extensions}.

A \emph{node} $j\in \NN$ may or may not have been provided with a replica of the database.
Nodes equipped with a replica ({\em i.e.} servers) can process requests from clients in their subtree. In other
words, there is a unique path from a client $i$ to the root of the tree, and each node in this path
is eligible to process some or all the requests issued by $i$ when provided with a replica.

Let $r$ be the root of the tree.
If $j\in \NN$, then $\child(j)$ is the set of children of node~$j$.
If $k \neq r$ is any node in the tree (leaf or internal), $\parent(k)$ is its parent in the tree.
If $l: k \rightarrow k'=\parent(k)$ is any link in the tree, then
$\suc(l)$ is the link $k' \rightarrow \parent(k')$ (when it exists).
Let $\Ancests(k)$ denote the set of ancestors of node $k$, i.e. the
nodes in the unique path that leads from $k$ up to the root $r$
($k$~excluded).
If $k' \in \Ancests(k)$, then
$\pat{k}{k'}$ denotes the set of links in the path from $k$ to $k'$; also,
$\subtree(k)$ is the subtree rooted in $k$, including~$k$.

We introduce more notations to describe our system in the following.
\begin{itemize}
\item {\bf Clients $i\in \CC$ --} Each client $i$ (leaf of the tree) is
sending $\rr_i$ requests per time unit.
For such requests, the required QoS (typically, a response time) is
denoted $\qos_i$, and we need to ensure that this QoS will be
satisfied for each client.
\item {\bf Nodes $j\in \NN$ --} Each node $j$ (internal node of the
tree) has a processing capacity $\W_j$, which is the total number of requests
that it can process per time-unit when it has a replica. A cost is also
associated to each node, $\sto_j$, which represents the price to pay
to place a replica at this node. With a single object type it is quite natural
to assume that $\sto_j$ is proportional to $\W_j$: the more powerful a server, the more
costly. But with several objects we may use non-related values of capacity and cost.
\item {\bf Communication links $l\in \LL$ --} The edges of the tree represent
the communication links between nodes (leaf and internal). We assign a
communication time $\comm_l$ on link $l$ which is the time required to send a request through the link.
Moreover, $\CL_l$ is the maximum number of requests that link $l$ can transmit
per time unit.
\end{itemize}


\subsection{Problem instances}

For each client $i \in \CC$, let $\Servers(i) \subseteq \NN$
be the set of servers responsible for processing at least one of its requests.
We do not specify here which access policy is enforced (e.g. one or multiple servers),
we defer this to Section~\ref{sec.strategies}. Instead, we let $\rr_{i,s}$ be the
number of requests from client $i$ processed by server $s$ (of course,
$\sum_{s \in \Servers(i)} \rr_{i,s} = \rr_i$).
In the following, $R$ is the set of replicas:
$$R = \left\{ s \in \NN | \; \exists i\in C \; , \;  s \in \Servers(i) \right\}.$$


\subsubsection{Constraints}

Three main types of constraints are considered.

\begin{description}

\item[Server capacity --] The constraint that no server capacity can be exceeded is present
in all variants of the problem:
$$ \forall s \in R, \sum_{i\in \CC | s \in \Servers(i)} \rr_{i,s} \leq \W_s$$

\item[QoS --] Some problem instances enforce a quality of service: the time to transfer a request
from a client to a replica server is bounded by a quantity $\qos_i$. This translates into:
$$ \forall i\in \CC, \forall s \in \Servers(i), \sum_{l \in \pat{i}{s}}  \comm_l \leq \qos_i.$$

Note that it would be easy to extend the QoS constraint so as to take
the computation cost of a request in addition to its communication
cost. This former cost is directly related to the computational
speed of the server and the amount of computation (in flops)
required for each request.

\item[Link capacity --] Some problem instances enforce a global
constraint on each communication link $l\in \LL$:
$$\sum_{i\in \CC, s \in \Servers(i) | l \in \pat{i}{s}} \rr_{i,s} \leq \BW_l $$

\end{description}


\subsubsection{Objective function}

The objective function for the \REP problem is defined as:
$$\text{Min} \sum_{s \in R} \sto_s$$

As already pointed out, it is frequently assumed that the cost of a server is
proportional to its capacity, so in some problem instances we let $\sto_s = \W_s$.

\subsubsection{Simplified problems}
\label{sec.simple}

We define a few simplified problem instances in the following:

\begin{description}

%

\item[QoS=distance --]
We can simplify the expression of the communication time in the QoS constraint
and only consider the distance (in number of hops) between a client
and its server(s). The QoS constraint is then
$$\forall i\in \CC, \forall s \in \Servers(i),\;  d(i,s) \leq \qos_i$$
where the distance $d(i,s)=|\pat{i}{s}|$ is the number of communication
links between $i$ and $s$.

\item[No QoS --]
We may further simplify the problem, by completely suppressing the
QoS constraints. In this case, the servers can be anywhere in the
tree, their location is indifferent to the client.

\item[No link capacity --] We may consider the problem assuming infinite link
capacity, i.e. not bounding the total traffic on any link in an admissible solution.

\item[Only server capacities --] The problem without QoS and link capacities
reduces to finding a valid solution of minimal cost, where ``valid'' means that
no server capacity is exceeded. We name \COR this fundamental problem.

\item[Replica counting --]
We can further simplify the previous \COR problem in the homogeneous case:
with identical servers, the \COR problem amounts to minimize the
number of replicas needed to solve the problem. In this case, the
storage cost $\sto_j$ is set to $1$ for each node. We call this
problem \RCO.

%

\end{description}

%
%
%
%
%
%
%

\section{Access policies}
\label{sec.strategies}



In this section we review the usual policies enforcing which replica
is accessed by a given client.
Consider that each client $i$ is
making $r_i$ requests per time-unit. There are two scenarios for the
number of servers assigned to each client:

\begin{description}
  \item[Single server --] Each client $i$ is assigned a single server
$\server(i)$, that is responsible for processing all its requests.
  \item[Multiple servers --] A client $i$ may be assigned several
servers in a set $\Servers(i)$. Each server $s \in \Servers(i)$ will
handle a fraction $\rr_{i,s}$ of the requests. Of course $\sum_{s \in
\Servers(i)} \rr_{i,s} = \rr_i$.
\end{description}

To the best of our knowledge, the single server policy has been
enforced in all previous approaches. One objective of this paper is to
assess the impact of this restriction on the performance of data
replication algorithms. The single
server policy may prove a useful simplification, but may come at the price
of a non-optimal resource usage.

In the literature, the single server strategy
 is further constrained to the \CLOSEST policy.
Here, the server of client $i$ is constrained to
be the first server found on the path that goes from $i$ upwards to
the root of the tree. In particular, consider a client $i$ and its server
$\server(i)$. Then any other client node $i'$ residing in the subtree
rooted in $\server(i)$ will be assigned a server in that subtree.
This forbids requests from $i'$ to ``traverse'' $\server(i)$
and be served higher (closer to the root in the tree).

We relax this constraint in the \UPWARDS policy which is the general
single server policy. Notice that a solution to \CLOSEST always is a
solution to \UPWARDS, thus \UPWARDS is always better than \CLOSEST
in terms of the objective function.
Similarly, the \MULTIPLE policy is always better than \UPWARDS, because  it is not
constrained by the single server restriction.

The following sections illustrate the three policies.
Section~\ref{sec.impact} provides simple examples where there is a valid solution
for a given policy, but none for a more constrained one. Section~\ref{sec.up-clo}
shows that \UPWARDS can be arbitrarily better than \CLOSEST, while
Section~\ref{sec.mul-up} shows that \MULTIPLE can be arbitrarily better than \UPWARDS.
We conclude with an example showing that the cost of an optimal solution of the \RCO problem
(for any policy) can be arbitrarily higher than the obvious lower bound
$$\left\lceil \frac{\sum_{i \in \CC} r_i}{W} \right\rceil,$$
where $W$ is the server capacity.



\subsection{Impact of the access policy on the existence of a solution}
\label{sec.impact}

\begin{figure}[htb]
\begin{center}
\includegraphics[width=0.9\textwidth]{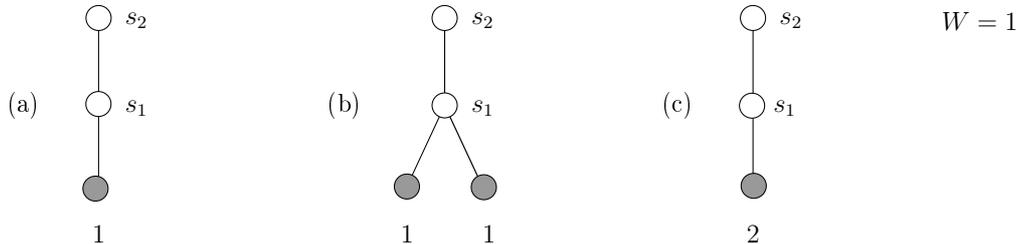}
\end{center}
\vspace{-.1cm}
\caption{Access policies.}
\label{fig.policy1}
\end{figure}

We consider here a very simple instance of the \RCO problem. In this example
there are two nodes, $s_1$ being the unique
child of $s_2$, the tree root (see Figure~\ref{fig.policy1}).
Each node can process $\W=1$ request.

\begin{compactitem}
\item If $s_1$ has one client child making $1$ request, the problem has a
solution with all three policies, placing a replica on $s_1$ or on $s_2$
indifferently (Figure~\ref{fig.policy1}(a)).
\item If $s_1$ has two client children, each making $1$ request, the problem
has no more solution with \CLOSEST. However, we have a solution with
both \UPWARDS and \MULTIPLE if we place replicas on both nodes. Each
server will process the request of one of the clients
(Figure~\ref{fig.policy1}(b)).
\item Finally, if $s_1$ has only one client child making $2$ requests,
only \MULTIPLE has a solution since we need to process one request on
$s_1$ and the other on $s_2$, thus requesting multiple servers
(Figure~\ref{fig.policy1}(c)).
\end{compactitem}

This example demonstrates the usefulness of the new policies. The \UPWARDS
policy allows to find solutions when the classical
\CLOSEST  policy does not. The same holds true for \MULTIPLE versus \UPWARDS.
In the following, we compare the cost of
solutions obtained with different strategies.

\subsection{\UPWARDS versus \CLOSEST}
\label{sec.up-clo}

In the following example, we construct an instance of \RCO where
the cost of the \UPWARDS policy is arbitrarily lower than the cost of the \CLOSEST
policy. We consider the tree network of Figure~\ref{fig.policy2}, where there
are $2n+2$ internal nodes, each with $\W_j = W = n$, and $2n+1$ clients, each with $\rr_i = r = 1$.

\begin{figure}[htb]
\begin{center}
\includegraphics{policy2.fig}
\end{center}
\vspace{-.1cm}
\caption{\UPWARDS versus \CLOSEST}
\label{fig.policy2}
\end{figure}

With the \UPWARDS policy, we place three replicas in $s_{2n}$, $s_{2n+1}$
and $s_{2n+2}$. All requests can be satisfied with these three replicas.

When considering the \CLOSEST policy, first we need to place a replica
in $s_{2n+2}$ to cover its client. Then,
\begin{compactitem}
\item Either we place a replica on $s_{2n+1}$. In this case, this
replica is handling $n$ requests, but there remain $n$ other requests from the $2n$ clients
in its subtree that 
cannot be processed by  $s_{2n+2}$. Thus, we need to add
$n$~replicas between $s_1..s_{2n}$.
\item Otherwise, $n-1$ requests of the $2n$ clients in the subtree of
$s_{2n+1}$ can be processed by $s_{2n+2}$ in addition to its own client. We need to
add
$n+1$~extra replicas among $s_1, s_2, \ldots, s_{2n}$.
\end{compactitem}
In both cases, we are placing $n+2$ replicas, instead of the $3$
replicas needed with the \UPWARDS policy.
This proves that \UPWARDS can be arbitrary better than \CLOSEST
on some \RCO instances.

\subsection{\MULTIPLE versus \UPWARDS}
\label{sec.mul-up}

In this section we build an instance of the \RCO problem where \MULTIPLE is twice better
than \UPWARDS. We do not know whether there exist instances of \RCO where the
performance ratio of \MULTIPLE versus \UPWARDS is higher than $2$
(and we conjecture that this is not the case).
However, we also build an instance of the \COR problem (with heterogeneous nodes)
where \MULTIPLE is arbitrarily better
than \UPWARDS.

\begin{figure}[htb]
\begin{center}
\includegraphics{policy3.fig}
\end{center}
\vspace{-.1cm}
\caption{\MULTIPLE versus \UPWARDS, homogeneous platforms.}
\label{fig.policy3}
\end{figure}

We start with the homogeneous case. Consider the instance of \RCO represented in
Figure~\ref{fig.policy3}, with $3n+1$ nodes of capacity $\W_j = W = 2n$. The root $r$
has $n+1$ children, $n$ nodes labeled $s_1$ to $s_n$ and a client with $\rr_i=n$.
Each node $s_j$ has two children nodes, labeled $v_j$ and $w_j$ for $1 \leq j \leq n$.
Each node $v_j$ has a unique child, a client with $\rr_i =n$ requests;  each node $w_j$ has
a unique child, a client with $\rr_i =n+1$ requests.

The \MULTIPLE policy assigns $n+1$ replicas, one to the root $r$ and one to each node $s_j$.
The replica in $s_j$ can process all the $2n+1$ requests in its subtree except one, which
is processed by the root.

For the \UPWARDS policy, we need to assign one replica to $r$, to cover its client. This replica
can process $n$ other requests, for instance those from the client child of $v_1$. We need to place
at least a replica in $s_1$ or in $w_1$, and $2(n-1)$ replicas in $v_j$ and $w_j$ for $2 \leq j \leq n$.
This leads to a total of $2n$ replicas, hence a performance factor $\frac{2n}{n+1}$
whose limit is to $2$ when $n$ tends to infinity.

\begin{figure}[htb]
\begin{center}
\includegraphics{policy4.fig}
\end{center}
\vspace{-.1cm}
\caption{\MULTIPLE versus \UPWARDS, heterogeneous platforms.}
\label{fig.policy4}
\end{figure}

\medskip
We now proceed to the heterogeneous case. Consider the instance of \COR represented in
Figure~\ref{fig.policy4}, with $3$ nodes $s_1$, $s_2$ and $s_3$, and $2$ clients.
The capacity of $s_1$ and $s_2$
is $\W_1 = \W_2 = n$ while that of $s_3$ is $\W_3 = Kn$, where $K$ is arbitrarily large.
Recall that in the \COR problem, we let $\sto_j = \W_j$ for each node.
\MULTIPLE assigns $2$ replicas, in $s_1$ and $s_2$, hence has cost $2n$.
The \UPWARDS policy assigns a replica to $s_1$ to cover its child, and then cannot use $s_2$
to process the requests of the child in its subtree. It must place a replica in $s_3$,
hence a final cost $n+Kn = (K+1)n$ arbitrarily higher than \MULTIPLE.

%
%

\subsection{Lower bound for the \RCO problem}
\label{sec.bound}

Obviously, the cost of an optimal solution of the \RCO problem
(for any policy) cannot be lower than the obvious lower bound
$\left\lceil \frac{\sum_{i \in \CC} r_i}{W} \right\rceil$, where $W$ is the server capacity.
Indeed, this corresponds to a solution where the total  request load is shared as evenly as possible
among the replicas.

\begin{figure}[htb]
\begin{center}
\includegraphics{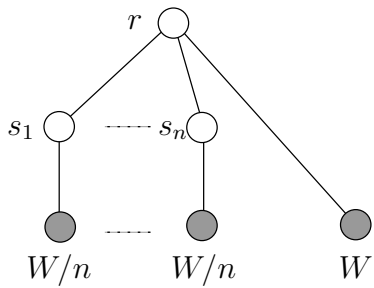}
\end{center}
\vspace{-.1cm}
\caption{The lower bound cannot be approximated for \RCO.}
\label{fig.lowerbound}
\end{figure}

The following instance of \RCO  shows that the optimal cost can be arbitrarily higher than this
lower bound. Consider Figure~\ref{fig.lowerbound}, with $n+1$ nodes of capacity $\W_j = W$,
The root $r$ has $n+1$ children, $n$ nodes labeled $s_1$ to $s_n$, and a client with $r_i= W$.
Each node $s_j$ has a unique child, a client with $r_i = W/n$ (assume without loss of generality
that $W$ is divisible by $n$). The lower bound is $\left\lceil
\frac{\sum_{i \in \CC} r_i}{W} \right\rceil = \frac{2W}{W} = 2$.
However, each of the three policies \CLOSEST, \UPWARDS and \MULTIPLE will assign a replica to the root
to cover its client, and will then need $n$ extra replicas, one per client of $s_j$, $1 \leq j \leq n$.
The total cost is thus $n+1$ replicas, arbitrarily higher than the lower bound.

All the examples in Sections~\ref{sec.impact} to~\ref{sec.bound} give an insight of the
combinatorial nature of the \REP optimization problem, even in its simplest variants
\COR and \RCO. The following section corroborates this insight: most problems are shown NP-hard, even though
some variants have polynomial complexity.

\section{Complexity results}
\label{sec.complexity}

One major goal of this paper is to assess the impact of the access policy
on the problem with homogeneous vs heterogeneous servers.
We restrict to the simplest problem, namely the \COR problem introduced in Section~\ref{sec.simple}.
We consider a tree $\TT = \CC \cup \NN$,
no QoS constraint, and infinite link capacities.
Each client $i \in \CC$ has $\rr_i$ requests; each node $j \in \NN$ has processing capacity $\W_j$
and storage cost $\sto_j = W_j$.
This simple problem comes in two flavors,
either with homogeneous nodes ($\W_j = \W$ for all $j \in \NN$), or with heterogeneous nodes
(servers with different capacities/costs).

In the single server version of the problem, we need to find a server $\server(i)$ for each client
$i \in \CC$. Let $\Servers$ be the set of servers chosen among the nodes in $\NN$. The only constraint
is that server capacities cannot be exceeded: this translates into
$$\sum_{i \in \CC, \server(i) =j} r_i \leq W_j \quad \text{ for all } j \in \Servers.$$
The objective is to find a valid solution of minimal storage cost $\sum_{j \in \Servers} W_j$.
Note that with homogeneous nodes, the problem reduces to find the minimum number of servers,
i.e. to the \RCO problem.
As outlined in Section~\ref{sec.strategies},
there are two variants of the single server version of the problem, namely the
\CLOSEST and the \UPWARDS strategies.

In the \MULTIPLE policy with multiple servers per client, let $\Servers$ be the set of servers chosen
among the nodes in $\NN$; for any client $i \in \CC$ and any node $j \in \NN$,
let $\rr_{i,j}$ be the number of requests from $i$ that are processed by $j$ ($\rr_{i,j}= 0$
if $j\notin \Servers$). We need to ensure
that
$$\sum_{j \in \NN} r_{i,j} = r_i \quad \text{ for all } i \in \CC.$$
The capacity constraint
now writes
$$\sum_{i \in \CC} r_{i,j} \leq W_j \quad \text{ for all } j \in \Servers,$$
while the objective function is the same as for the single server version.

The decision problems associated with the previous optimization problems are
easy to formulate: given a bound on the number of servers (homogeneous version)
or on the total storage cost (heterogeneous version), is there a valid solution
that meets the bound?

\begin{table}
  \centering
\begin{tabular}{|l|l|l|}
  \hline
   & \textbf{Homogeneous} & \textbf{Heterogeneous} \\
  \textbf{\CLOSEST} & polynomial~\cite{Cidon2002,PangfengLiu06} & NP-complete \\
  \textbf{\UPWARDS} & NP-complete &  NP-complete\\
  \textbf{\MULTIPLE} & polynomial & NP-complete \\
  \hline
\end{tabular}
  \caption{Complexity results for the different instances of the \COR problem.}
  \label{tab.summary}
\end{table}

Table~\ref{tab.summary} captures the complexity results. These
complexity results
are all new, except for the \CLOSEST/Homogeneous combination.
The NP-completeness of the \UPWARDS/Homogeneous case comes as a
surprise, since all
previously known instances 
were shown to be polynomial, using dynamic programming
algorithms. In particular,
the \CLOSEST/Homogeneous variant remains polynomial when adding
communication costs~\cite{Cidon2002} or QoS
constraints~\cite{PangfengLiu06}. Previous NP-completeness results
involved general graphs rather than trees, and the combinatorial
nature of the problem came from the difficulty to extract a good
replica tree out of an arbitrary communication graph. Here the tree is
fixed, but the problem remains combinatorial due to resource
heterogeneity.

\subsection{With homogeneous nodes and the \MULTIPLE strategy}
\label{sec.elegant}

\begin{theorem}
The instance of the \RCO problem with the \MULTIPLE strategy
can be solved in polynomial time.
\end{theorem}

\begin{proof}
We outline below an optimal algorithm to solve the problem.
The proof of optimality is quite technical, so the reader may want to
skip it at first reading.
\end{proof}

\subsubsection{Algorithm for multiple servers}

We propose a greedy algorithm to solve the \RCO problem.
Let $\W$ be the total number of requests that a server can handle.

This algorithm works in three passes: first we select the nodes which
will have a replica handling exactly $\W$ requests.
Then a second pass allows us to select some
extra servers which are fulfilling the remaining requests.
Finally, we need to decide for each server how many requests of each
client it is processing.

We assume that each node $i$ knows its parent $\parent(i)$ and its
children $\child(i)$ in the tree.
We introduce a new variable which is the flow coming up in the tree
(requests which are not already fulfilled by a server). It is denoted
by $\flow_i$ for the flow between $i$ and $\parent(i)$.
Initially, $\forall i\in \CC \; \flow_i=\rr_i$ and
$\forall i\in \NN \; \flow_i=-1$.
Moreover, the set of replicas is empty in the beginning: $repl=\emptyset$.

\begin{description}
\item[Pass 1--] We greedily select in this step some nodes which will process
$\W$ requests and which are as close to the leaves as possible. We place
a replica on such nodes (see Algorithm~\ref{alg:pass1}). Procedure
{\bf pass1} is called with $r$ (root of the tree) as a parameter,
and it goes down the
tree recursively in order to compute the flows. When a flow exceeds
$\W$, we place a replica since the corresponding server will be fully
used, and we remove the processed requests from the flow going upwards.

At the end, if $flow_r =0$ or ($flow_r \leq \W$ and $r \notin repl$),
we have an optimal  solution since all
replicas which have been placed are
fully used and all requests are satisfied by adding a replica in $r$
if $flow_r \neq 0$. In this case we skip pass~2 and go directly to pass~3.

Otherwise, we need some extra replicas since some requests are not
satisfied yet, and the root cannot satisfy all the remaining requests.
To place these extra replicas, we go through pass~2.

\begin{algorithm}[bthp]
\label{alg:pass1}
\SetLine
\caption{Procedure pass1}
procedure {\bf pass1} (node $s\in \NN$)\\
\Begin{
  $flow_s = 0$\;
  \For{$i \in \child(s)$}{
     \lIf{$flow_i==-1$}{{\bf pass1}$(i)$; {\em // Recursive call.}\\}
     $flow_s = flow_s + flow_i$\;
  }
  \lIf{$flow_s \geq \W$}{$flow_s = flow_s-\W; repl= \{s\} \cup repl$\;}
}
\end{algorithm}

\item[Pass 2--]
In this pass, we need to select the nodes where to add replicas.
To do so, while there are too many requests going up to the root, we
select the node which can process the highest number of requests,
and we place a replica there.
The number of requests that a node $j\in \NN$ can eventually process is the
minimum of the flows between $j$ and the root $r$, denoted $uflow_j$
(for {\em u}seful {\em flow}). Indeed, some
requests may have no server yet, but they might be processed by a
server on the path between $j$ and $r$, where a replica has been
placed in pass~1.
Algorithm~\ref{alg:pass2} details this pass.

If we exit this pass with $finish=-1$, this means that we have tried
to place replicas on all nodes, but this solution is not feasible
since there are still some requests which are not processed going up
to the root. In this case, the original problem instance had no solution.

However, if we succeed to place replicas such that $flow_r = 0$, 
we have a set of replicas which succeed to process all requests. We
then go through pass~3 to assign requests to servers, i.e. to compute how many requests of
each client should be processed by each server.

\begin{algorithm}[tbhp]
\label{alg:pass2}
\SetLine
\caption{Pass 2}
\While{$flow_r \neq 0$}{
  $freenode=\NN \setminus repl$\;
  \lIf{$freenode==\emptyset$}{$finish=-1;$ exit the loop\;}
  {\em // At each step, assign 1 replica and re-compute flows.}\\
  $child=\child(r); uflow_r = flow_r$\;
  \While{$child != \emptyset$}{
    remove $j$ from $child$\;
    $uflow_j = \min(flow_j, uflow_{\parent(j)})$\;
    $child = child \cup \child(j)$\;
  }
  {\em // The useful flows have been computed, select the max.}\\
  maxuflow=0\;
  \For{$j\in freenode$}{
   \lIf{$uflow_j > maxuflow$}{$maxuflow=uflow_j;$ $maxnode=j$\;} }
  \If{$maxuflow\neq 0$} {$repl = repl \cup \{maxnode \}$\;
    {\em // Update the flows upwards.}\\
    \lFor{$j \in \Ancests(maxnode)\cup \{maxnode\}$}
       {$flow_j = flow_j-maxuflow$\;}
  }
  \lElse{$finish=-1;$ exit the loop\;}
}
\end{algorithm}

\item[Pass 3--]
This pass is in fact straightforward, starting from the leaves and
distributing the requests to the servers from the bottom until the top
of the tree. We decide for instance to affect requests from clients
starting to the left. Procedure
{\bf pass3} is called with $r$ (root of the tree) as a parameter,
and it goes down the tree recursively (c.f. Algorithm~\ref{alg:pass3}).
For $i\in \CC$, $r'_i$ is the
number of requests of $i$ not yet affected to a server (initially
$r'_i=\rr_i$). $w_{s,i}$ is the number of requests of client~$i$
affected to server $s \in \NN$, and $w_s\leq \W$ is the total number
of requests affected to $s$.
$C(s)$ is the set of clients in $\subtree(s)$ which still have some
requests not affected. Initially, $C(i)= \{i\}$ for $i\in \CC$, and
$C(s)=\emptyset$ otherwise.

Note that a server which was computing $\W$ requests
in pass~1 may end up computing fewer requests if one of its descendants in the
tree has earned a replica in pass~2. But this does not affect the
optimality of the result, since we keep the same number of replicas.
\end{description}

\begin{algorithm}[bthp]
\label{alg:pass3}
\SetLine
\caption{Procedure pass3}
procedure {\bf pass3} (node $s\in \NN$)\\
\Begin{
  $w_s=0$\;
  \For{$i \in \child(s)$}{
     \lIf{$C(i)=\emptyset$}{{\bf pass3}$(i)$; {\em // Recursive call.}\\}
     $C(s) = C(s) \cup C(i)$\;
  }
  \If {$s\in repl$} {\For{$i \in C(s)$}
     {\lIf{$r'(i)\leq \W - w_s$}
        {$C(s)=C(s)\setminus \{i\};$ $w_{s,i}=r'_i$; $w_s=w_s+r'_i;$
$r'_i =0$\;}}
     \lIf{$C(s)\neq \emptyset$} {Let $i\in C(s);$ $x=\W-w_s;$
        $r'_i=r'_i-x;$ $w_{s,i}=x;$ $w_s=\W$\;}
   }
}
\end{algorithm}

The proof in Section~\ref{sec:mul-proof} shows the equivalence between
the solution built by this algorithm and any optimal solution, thus
proving the optimality of the algorithm. The following example
illustrates the step by step execution of the algorithm.

\subsubsection{Example}

Figure~\ref{fig.exalgo}(a) provides an example of network on which we
are placing replicas with the \MULTIPLE strategy. The network is thus
homogeneous and we fix $\W=10$.

\begin{figure}[Htb]
\begin{center}
\includegraphics[width=15cm]{exalgo.fig}
\end{center}
\vspace{-.1cm}
\caption{Algorithm for the \RCO problem with the \MULTIPLE strategy.}
\label{fig.exalgo}
\end{figure}

Pass~1 of the algorithm is quite straightforward to unroll, and
Figure~\ref{fig.exalgo}(b) indicates the flow on each link
and the saturated replicas are the black nodes.

During pass~2, we select the nodes of maximum useful flow.
Figure~\ref{fig.exalgo}(c) represents these useful flows; we see
that node~$n_4$ is the one with the maximum useful flow~($7$), so we
assign it a replica and update the useful flows. All the useful
flows are then reduced down to $1$ since there is only $1$ request
going through the root $n_1$. The first node of maximum useful
flow~$1$ to be selected is $n_2$, which is set to be a replica of
pass~2. The flow at the root is then~$0$ and it is the end of
pass~2.

Finally, pass~3 affects the servers to the clients and decides which
requests are  served by which replica (Figure~\ref{fig.exalgo}(d)).
For instance, the client with $12$ requests shares its requests
between $n_{10}$ ($10$ requests) and $n_2$ ($2$ requests). Requests
are affected from the bottom of the tree up to the top. Note that
the root $n_1$, even though it was a saturated replica of pass~1,
has only $5$ requests to proceed in the end.


\subsubsection{Proof of optimality}
\label{sec:mul-proof}

Let $R_{opt}$ be an optimal solution to an instance of the
problem. The core of the proof consists in transforming this
solution into an equivalent canonical optimal solution $R_{can}$. We
will then show that our algorithm is building this canonical solution,
and thus it is producing an optimal solution.

Each server $s\in R_{opt}$ is serving $w_{s,i}$ requests of client
$i\in \subtree(s) \cap \CC$, and
$$w_s = \sum_{i\in \subtree(s)\cap \CC} w_{s,i} \leq W.$$
For each $i\in \CC$, $w_{s,i}=0$ if $s\in\NN$ is not a replica,
and, $\sum_{s\in Ancests(i)} w_{s,i} = \rr_i$.

We define the {\em flow} of node $k$, $\flow_k$, by the number of requests
going through this node up to its parents. Thus, for $i\in \CC$,
$flow_i = \rr_i$, while for a node $s\in \NN$,
$$flow_s = \sum_{i\in \child(s)} flow_i - w_s.$$

The {\em total flow} going through the tree, $tflow$, is defined in a similar
way, except that we do not remove from the flow the requests processed
by a replica, {\em i.e.} $tflow_s = \sum_{i\in \child(s)} tflow_i$.
We thus have $$tflow_s = \sum_{i \in \subtree(s) \cap \CC} \rr_i.$$


These variables are completely defined by the network and the optimal
solution $R_{opt}$.

A first lemma shows that it is possible to change request
assignments while keeping an optimal solution. The flows need to be
recomputed after any such modification.

\begin{lemma}
\label{req-aff}
Let $s \in \NN \cap R_{opt}$ be a server such that $w_s < \W$.
\begin{compactitem}
\item If $tflow_s \geq \W$, we can change the request assignment
between replicas of the optimal solution, in such a way that $w_s = \W$.
\item Otherwise, we can change the request assignment so that
$w_s = tflow_s$.
\end{compactitem}
\end{lemma}
\begin{proof}
First we point out that the clients in $\subtree(s)$ can
all be served by $s$, and since $R_{opt}$ is a solution, these
requests are served by a replica somewhere in the tree. We do not
modify the optimality of the solution by changing the $w_{s,i}$, it
just affects the flows of the solution. Thus, for a given client
$i\in \subtree(s)\cap\CC$, if there is a replica $s'\neq s$ on the
path between $i$ and the root, we can change the assignment of the
requests of client $i$. Let $x = max (w_{s',i}, \W-w_s)$. Then we
move $x$ requests, {\em i.e.} $w_{s',i} = w_{s',i} - x$ and $w_{s,i} =
w_{s,i} + x$. From the definition of $tflow_s$, we obtain the result,
if we move all possible requests to $s$ until there are no more
requests in the subtree or until $s$ is processing $W$ requests.
\end{proof}

We now introduce a new definition, completely independent from the
optimal solution but related to the tree network.
The {\em canonical flow} is obtained by distinguishing nodes which
receive a
flow greater than $W$ from the other nodes. We compute the canonical
flow $cflow$ of the tree, independently of the replica placement, and define
a subset of nodes which are {\em saturated}, $SN$.
We also compute the number of saturated nodes in $\subtree(k)$,
denoted $nsn_k$, for any node $k \in \CC \cup \NN$ of the tree.

For $i \in \CC$, $cflow_i = r_i$ and $nsn_i=0$, and we then compute
recursively the canonical flows for nodes $s\in \NN$.
Let $f_s = \sum_{i\in \child(s)} cflow_i$ and
$x_s =  \sum_{i\in \child(s)} nsn_i$.
If $f_s \geq \W$ then $s \in SN$,
$cflow_s = f_s - \W$ and $nsn_s=x_s + 1$.
Otherwise, $s$ is not saturated, $cflow_s = f_s$ and $nsn_s=x_s$.

We can deduce from these definitions the following results:
\begin{proposition}
\label{prop1}
A non saturated node always has a canonical flow being less than $\W$:\\
$\forall s \in \NN \setminus SN \;\; cflow_s < \W$
\end{proposition}

\begin{lemma}
\label{prop2a}
For all nodes $s \in \CC \cup \NN$, 
$cflow_s = tflow_s - nsn_s \times \W$.
\end{lemma}

\begin{corollary}
\label{prop2}
For all nodes $s \in \CC \cup \NN$, 
$tflow_s \geq nsn_s \times \W$.
\end{corollary}

\begin{proof}
Proposition~\ref{prop1} is trivial due to the definition of the
canonical flow.

Lemma~\ref{prop2a} can be proved recursively on the tree.
\begin{compactitem}
\item This property is true for the clients:
for $i\in \CC$, $nsn_i = 0$ and $tflow_i = cflow_i = \rr_i$.
\item Let $s\in \NN$, and let us assume that the proposition is true for all
children of $s$. Then,
$$\forall j\in \child(s) \; cflow_j = tflow_j - nsn_j \times \W.$$
\begin{compactitem}
\item If $s \notin SN$, $nsn_s = \sum_{j\in \child(s)} nsn_j$ and
$$cflow_s = \sum_{j\in \child(s)} cflow_j
 =  \sum_{j\in \child(s)} (tflow_j - nsn_j \times \W)
 =  tflow_s - nsn_s \times \W $$
\item If $s\in SN$, $nsn_s =  \left(\sum_{j\in \child(s)} nsn_j\right) + 1$ and
$$cflow_s = \sum_{j\in \child(s)} cflow_j - \W
 =  \sum_{j\in \child(s)} (tflow_j - nsn_j \times \W) - \W$$
$$ =  tflow_s - (nsn_s - 1) \times \W - \W
 =  tflow_s - nsn_s \times \W$$
\end{compactitem}
\end{compactitem}
which proves the result.
Corollary~\ref{prop2} is trivially deduced from
Lemma~\ref{prop2a} since $cflow$ is a positive function.
\end{proof}

We also show that it is always possible to move a replica into a free
server which is one of its ancestors in the tree, while keeping an optimal
solution:

\begin{proposition}
\label{repl-up}
Let $R_{opt}$ be an optimal solution, and let $s\in R_{opt}$.
If $\exists s' \in \Ancests(s) \setminus R_{opt}$ then
$R_{opt}' = \{s'\} \cup R_{opt}\setminus \{s\}$ is also an optimal
solution.
\end{proposition}
\begin{proof}
$s'$ can handle all requests which
were processed by $s$ since $s \in \subtree(s')$. We just need to
redefine $w_{s',i} = w_{s,i}$ for all $i\in \CC$ and then $w_{s,i}=0$.
\end{proof}

We are now ready to transform $R_{opt}$ into a new optimal solution,
$R_{sat}$, by redistributing the
requests among the replicas and moving some replicas, in order to
place a replica at each saturated node, and affecting $\W$ requests to
this replica.
This transformation is done starting at the leaves of the tree,
and considering all nodes of $SN$.
Nothing needs to be done for the leaves (the clients) since they are
not in $SN$.

Let us consider $s\in SN$, and assume that the optimal solution
has already been modified to place a replica,  and assign it $\W$
requests,  on all nodes in $subSN = SN \cap \subtree(s) \setminus \{s\}$.

We need to differentiate two cases:
\begin{compactenum}
\item If $s \in R_{opt}$, we do not need to move any replica. However,
if $w_s \neq \W$, we change the assignment of some requests while
keeping the same replicas in order to obtain a workload of $\W$ on
server~$s$. We do not remove requests from the saturated servers
of $subSN$ which have already been filled.
Corollary~\ref{prop2}
ensures that $tflow_s \geq nsn_s \times \W$, and $(nsn_s-1)\times \W$
requests should not move since they are affected to the $nsn_s-1$
servers of $subSN$. There are thus still more than $\W$ requests of
clients of $\subtree(s)$ which can possibly be moved on $s$ using
Lemma~\ref{req-aff}.

\item If $s \notin R_{opt}$, we need to move a replica of $R_{opt}$
and place it in $s$ without changing the optimality of the
solution. We differentiate two subcases.
  \begin{compactenum}
  \item If $\exists s_1 \in \subtree(s) \cap R_{opt} \setminus SN$,
then the replica placed on $s_1$ can be moved in $s$ by applying
Proposition~\ref{repl-up}.
Then, if $w_s \neq \W$, we apply case~1 above to saturate the
server.

  \item Otherwise, all the replicas placed in $\subtree(s)$ are also
in $SN$, and the flow consumed by the already modified optimal
algorithm is exactly $(nsn_s-1) \times \W$.
It is easy to see that the flow (of the optimal solution) at $s$ is
exactly equal to the total flow minus the consumed flow. Therefore,
$flow_s = tflow_s - (nsn_s-1) \times \W$, and with the application of
Corollary~\ref{prop2}, $flow_s \geq \W$.

The idea now consists in affecting the requests of this flow to
node~$s$ by removing work from the replicas upwards to the root, and
rearrange the remaining requests to remove one replica.
The flow $flow_s$ is going upwards to be processed by some of the
$nr_s$ replicas in $\Ancests(s)\cap R_{opt}$,
denoted $s_1, ..., s_{nr_s}$, $s_1$ being the closest node from $s$.
We can remove $\W$ of these requests from the flow and affect them to
a new replica placed in $s$.
Let $w_{s_k,s}= \sum_{j \in \subtree(s) \cap \CC} w_{s_k,j}$.
We have
$\sum_{k=1..nr_s} w_{s_k, s} = flow_s$. We move these requests from
$s_k$ to $s$, starting with $k=1$. Thus, after the modification,
$w_{s_1, s} = 0$. It is however possible that $w_{s_1} \neq 0$ since
$s_1$ may process requests which are not coming from $\subtree(s)$.
In this case, we are sure that we have removed enough requests from
$s_k$, $k=2..nr_s$ which can instead process requests still in charge
of $s_1$. We can then remove the replica initially placed in $s_1$.

This way, we have not changed the assignment on replicas in $subSN$,
but we have placed a replica in $s$ which is processing $\W$
requests. Since we have at the same time removed the first replica
on the path from $s$ to the root ($s_1$), we have not changed the
number of replicas and the solution is still optimal.



  \end{compactenum}
\end{compactenum}

Once we have applied this procedure up to the root, we have an optimal
solution $R_{sat}$ in which all nodes of $SN$ have been placed a
replica and are processing $\W$ requests. We will not change the
assignment of these replicas anymore in the following.
Free nodes in the new solution are called {\em F-nodes}, while
replicas which are not in $SN$ are called {\em PS-nodes}, for
{\em partially saturated}.

In a next step, we further modify the $R_{sat}$ optimal solution in
order to obtain what we call the {\em canonical solution} $R_{can}$.
To do so, we change the request assignment of the PS-nodes: we
``saturate'' some of them as much as we can and we integrate them into
the subset of nodes $SN$, redefining the $cflow$ accordingly.
At the end of the process, $SN=R_{can}$.

The $cflow$ is still the flow which has not been processed by a
saturated node in the subtree, and thus we can express it in a more
general way:
$$cflow_s = tflow_s - \sum_{s'\in SN\cap \subtree(s)} w_{s'}$$
Note that this is totally equivalent to the previous definition
while we have not modified $SN$.

We also introduce a new flow definition, the {\em non-saturated flow}
of $s$,
$nsflow_s$, which counts the requests going through node~$s$ and not
served by a saturated server anywhere in the tree. Thus,
$$nsflow_s = cflow_s - \sum_{i\in \child(s)\cap \CC}
   \sum_{s'\in \Ancests(s) \cap SN}  w_{s',i}.$$
This flow represents the requests that can potentially be served
by~$s$ while keeping all nodes of SN saturated.


\begin{lemma}
\label{lemma:ps-ps}
In a saturated optimal solution, there cannot exist a PS-node in the
subtree of another PS-node.
\end{lemma}

\begin{proof}

The non-saturated flow is $nsflow_s \leq cflow_s$ since we
further remove from the canonical flow some requests which are
affected upwards in the tree to some saturated servers.

Let $s \in R_{sat} \setminus SN$ be a PS-node. Its canonical flow is
$cflow_s < W$. It can potentially process all the requests of the
subtree which are not affected to a saturated server upwards or
downwards in the tree, thus $nsflow_s$ requests.
Since $nsflow_s\leq cflow_s<W$, we can change the request assignment to
assign all these $nsflow_s$ requests to $s$, removing eventually some
work from other non-saturated replicas upwards or downwards which were
processing these requests.
Thus, the replica on node $s$ is processing all the
requests of $\subtree(s)$ which are not processed by saturated nodes.

If there was a non saturated replica in $\subtree(s)$, it could thus be
removed since all the requests are processed by $s$.
This means that a solution with a PS-node in the subtree of another
PS-node is not optimal, thus proving the lemma.
\end{proof}





At this point, we can move the PS-nodes as high as possible in
$R_{sat}$. Let $s$ be a PS-node. If there is a free node $s'$ in
$\Ancests(s)$ then we can move the replica from $s$ to $s'$ using
Proposition~\ref{repl-up}. Lemma~\ref{lemma:ps-ps} ensures that there are
no other PS-nodes in $\subtree(s')$.

All further modifications will only alter nodes which have no PS-nodes
in their ancestors. We define $\NNN= \{ s | \Ancests(s)\setminus SN
= \emptyset \}$.

Let $s\in \NNN$.
$nsflow_s = cflow_s - \sum_{i\in \child(s)\cap \CC}
   \sum_{s'\in \Ancests(s)}  w_{s',i}$ since all ancestors of $s$ are
in $SN$. Thus,
$$nsflow_s = \sum_{s' \in \subtree(s) \setminus SN} w_{s'}.$$

By definition, $\forall s\in \NN \; nsflow_s \leq cflow_s$.
Moreover, if $s \notin SN$, then $nsflow_s=w_s$ since
$\subtree(s)\setminus SN$ is reduced to $s$ (no other PS-node under
the PS-node~$s$, from Lemma~\ref{lemma:ps-ps}).

We introduce a new flow definition, the useful flow, which intuitively
represents the number of requests that can possibly be processed on
$s$ without removing requests from a saturated server.
$$uflow_s = \min_{s'\in \Ancests(s)\cup \{s\}} \{cflow_{s'} \}$$

\begin{lemma}
\label{nsflow:uflow}
Let $s\in \NNN$. Then $nsflow_s \leq uflow_s$.
\end{lemma}
\begin{proof}
Let $s' \in \Ancests(s)$. Since $s\in \NNN$, $s' \in SN$.
$$cflow_{s'} \geq nsflow_{s'} = \sum_{s'' \in \subtree(s') \setminus
SN} w_{s''}$$
But since $s\in \subtree(s')$, $\subtree(s)\setminus SN \subseteq
\subtree(s')\setminus SN$, hence $nsflow_{s}\leq
nsflow_{s'}$. Note that $nsflow$ is a non decreasing function (when
going up the tree).

Thus, $\forall s' \in \Ancests(s) \cup \{s\}$, $nsflow_s \leq cflow_{s'}$,
and by definition of the useful flow, $nsflow_s \leq uflow_s$.
\end{proof}

Now we start the modification of the optimal solution in order to
obtain the canonical solution. At each step, we select a node~$s\in
\NN\setminus SN$ maximizing the useful flow. If there are several
nodes of identical $uflow$, we select the first one in a depth-first
traversal of the tree. We will prove that we can affect $uflow_s$
requests to this node without unsaturating any server of SN. $s$ is
then considered as a saturated node, we recompute the canonical flows
(and thus the useful flows) and
reiterate the process until $cflow_r=0$, which means that all the
requests have been affected to saturated servers.

Let us explain how to reassign the requests in order to saturate $s$
with $uflow_s$ requests. The idea is to remove some requests from
$\Ancests(s)$ in order to saturate $s$, and then to saturate the
ancestors of $s$ again, by affecting them some requests coming from
other non saturated servers.

First, we note that $uflow_s \leq cflow_r = nsflow_r$. Thus,
$$uflow_s \leq \sum_{s'\in \NN \setminus SN} w_{s'} = w_s + \sum_{s'
\in PS}  w_{s'}$$
where PS is the set of non saturated nodes without~$s$.
Let $x=uflow_s-w_s$. If $x=0$, $s$ is already saturated. Otherwise, we
need to reassign $x$ requests to $s$. From the previous equation,
we can see that $\sum_{s' \in PS}  w_{s'} \geq uflow_s - w_s =
x$. There are thus enough requests handled by non saturated nodes
which can be passed to $s$.

The number of requests of $\subtree(s)\cap \CC$ handled by
$\Ancests(s)$ is
$$\sum_{s'\in \Ancests(s)} \sum_{i\in \subtree(s)\cap\CC} w_{s',i}= cflow_s-nsflow_s$$
by definition of the flow.
Or $cflow_s - nsflow_s \geq uflow_s - w_s = x$ so there are at least
$x$ requests that $s$ can take from its ancestors.

Let $a_1=\parent(s), ..., a_k=r$ be the ancestors of $s$.
$x_j = \sum_{i\in \subtree(s)\cap \CC} w_{a_j,i}$ is the amount of
requests that $s$ can take from $a_j$. We choose arbitrary where to
take the requests if $\sum_{j} x_j > x$, and do not modify the
assignment of the other requests. We thus assume in the following
that $\sum_j x_j =x$.
Since these $x_j$ requests are coming from a client in $\subtree(s)$,
we can assign them to $s$, and there are now only $\W - x_j$ requests
handled by $a_j$, which means that $a_j$ is temporarily unsaturated.
However, we have given $x$ extra requests to $s$, hence $s$ is
processing $w_s + x = uflow_s$ requests.

We finally need to reassign requests to $a_j, j=1..k$ in order to
saturate these nodes again, taking requests out of nodes in $PS$ (non
saturated nodes other than $s$). This is done iteratively starting
with $j=1$ and going up to the root $a_k$. At each step $j$, we assume
that $a_{j'}, j'<j$ have already been saturated again and we should
not move requests away from them. However, we can still eventually
take requests away from $a_{j''}, j''>j$.

In order to saturate $a_j$, we need to take:
\begin{compactitem}
\item either requests from $\subtree(a_j)\cap \CC$ which are currently
handled by $a_{j''}, j''>j$, but without moving requests which are
already affected to $s$ ({\em i.e.} $\sum_{j''>j} x_{j''}$);
\item or requests from non saturated servers in $\subtree(a_j)$,
except requests from $s$ and requests already given to $s$ that should
not be moved any more ({\em i.e.} $\sum_{j'<j} x_{j'}$).
\end{compactitem}
The number of requests that we can potentially affect to $a_j$ is
therefore:
$$X= \sum_{s' \in \subtree(a_j)\setminus SN \setminus \{s\}} w_{s'}
   + \sum_{i\in \subtree(a_j)\cap \CC} \sum_{s' \in \Ancests(a_j)} w_{s',i}
   - \sum_{j'<j}x_{j'} - \sum_{j''>j}x_{j''}$$

Let us show that $X \geq x_j$. Then we can use these requests to
saturate $a_j$ again.

$$cflow_{a_j} = nsflow_{a_j} + \sum_{i\in \subtree(a_j)\cap \CC}
\sum_{s'\in \Ancests(a_j)} w_{s',i}\\
    = w_s + X + \sum_{j'<j}x_{j'} + \sum_{j''>j}x_{j''}
    = X + w_s + x - x_j$$

But $cflow_{a_j} \geq uflow_s$ and $uflow_s - w_s = x$ so
$$ X = cflow_{a_j} - w_s -x + x_j \geq uflow_s - w_s - x + x_j = x_j$$

It is thus possible to saturate $s$ and then keep its ancestors
saturated. At this point, $s$ becomes a node of $SN$ and we can
recompute the canonical and non saturated flows. We have removed
$uflow_s$ requests which were processed by non saturated servers, so
the $cflow$ and $nsflow$ of all ancestors of $s$, including $s$,
should be decreased by $uflow_s$.

In particular, at the root, $cflow_r = cflow_r - uflow_s$, which
proves that the contribution of $s$ on $cflow_r$ is $uflow_s$.

In the last step of the proof, we show that the number of replicas in
the modified canonical solution at the end of the iteration
$R_{can}= SN$ has exactly the same number of replicas than $R_{sat}$.
In the saturated solution, each PS-node $s$ is processing $nsflow_s$
requests, while in the canonical solution, it is $uflow_s$. However,
at every step when adding a saturated node $s$, we have $uflow_s$
greater than any of the $nsflow$s. It is thus easy to see that the
number of nodes in the canonical solution is less or equal to the
number of nodes in the saturated solution. Since the saturated
solution is optimal, $|R_{can}|=|R_{sat}|$, which completes the proof.

Our algorithm builds $R_{can}$ in polynomial time, which
assesses the complexity of the problem.

\subsection{With homogeneous nodes and the \UPWARDS strategy}

\begin{theorem}
\label{th.up-homo}
The instance of the \RCO problem with the \UPWARDS strategy
is NP-complete in the strong sense.
\end{theorem}

\begin{proof}
The problem clearly belongs to
the class NP: given a solution, it is easy to verify in polynomial time that all requests are served
and that no server capacity is exceeded. To establish the completeness in the strong sense,
we use a reduction from
3-PARTITION~\cite{GareyJohnson}. We consider an instance $\II_1$ of 3-PARTITION:
given $3m$ positive integers $a_1, a_2, \ldots, a_{3m}$ such that
$B/4 < a_i < B/2$ for $1 \leq i \leq 3m$, and $\sum_{i=1}^{3m} a_i = mB$,
can we partition these integers into $m$ triples, each of sum $B$?
We build the following instance $\II_2$ of \RCO (see Figure~\ref{fig.proof1}):
\begin{itemize}
  \item $3m$ clients $c_i$ with $r_i = a_i$ for $1 \leq i \leq 3m$.
  \item $m$ internal nodes $n_j$ with $\W_j = \sto_j = B$ for $1 \leq j \leq m$.\\
  - The children of $n_1$ are all the $3m$ clients $c_i$, and its parent is $n_2$.\\
  - For $2 \leq j \leq m$, the only child of $n_j$ is $n_{j-1}$. For $1 \leq j \leq m-1$, the
  parent of $n_j$ is $n_{j+1}$ (hence $n_m$ is the root).
\end{itemize}
Finally, we ask whether there exists a solution with total storage cost $mB$, i.e. with a replica located at
each internal node.
Clearly, the size of $\II_2$ is polynomial (and even linear) in the size of $\II_1$.

\begin{figure}
  \centering
  \includegraphics{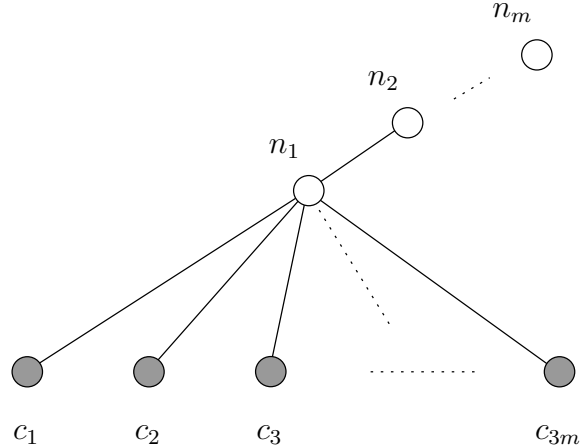}
  \caption{The platform used in the reduction for Theorem~\ref{th.up-homo}.}
  \label{fig.proof1}
\end{figure}

We now show that instance $\II_1$ has a solution if and only if instance $\II_2$ does.
Suppose first that $\II_1$ has a solution. Let $(a_{k_1},a_{k_2},a_{k_3})$ be the $k$-triplet in
$\II_1$. We assign the three clients $c_{k_1}$, $c_{k_2}$ and $c_{k_3}$ to server $n_k$.
Because $a_{k_1} + a_{k_2} + a_{k_3} = B$, no server capacity is exceeded. Because the $m$ triples
partition the $a_i$, all requests are satisfied. We do have a solution to $\II_2$.

Suppose now that $\II_2$ has a solution. Let $I_k$ be the set of clients served by node $n_k$
if there is a replica located at $n_k$: then $\sum_{i \in I_k} a_i \leq B$. The total number of
requests to be satisfied is $\sum_{i=1}^{3m} a_i = mB$, and there are
at most $m$ replicas of capacity $B$.
Hence no set $I_k$ can be empty, and $\sum_{i \in I_k} a_i \leq B$ for $1 \leq k \leq m$.
Because $B/4 < a_i < B/2$, each $I_k$ must be a triple. This leads to the desired solution of $\II_1$.
\end{proof}

\subsection{With heterogeneous nodes}

\begin{theorem}
\label{th.nphetero}
All three instances of the \COR problem with heterogeneous
nodes are NP-complete.
\end{theorem}

\begin{proof}
Obviously, the NP-completeness of the \UPWARDS strategy is a consequence
of Theorem~\ref{th.up-homo}. For the other two strategies, the problem clearly belongs to
the class NP: given a solution, it is easy to verify in polynomial time that all requests are served
and that no server capacity is exceeded. To establish the completeness, we use a reduction from
2-PARTITION~\cite{GareyJohnson}. We consider an instance $\II_1$ of 2-PARTITION:
given $m$ positive integers $a_1, a_2, \ldots, a_m$,
does there exist a subset $I \subset \{1, \ldots, m\}$
such that $\sum_{i \in I} a_i = \sum_{i \notin I} a_i$. Let $S = \sum_{i=1}^{m} a_i$.
We build the following instance $\II_2$ of \COR (see Figure~\ref{fig.proof2}):
\begin{itemize}
  \item $m+1$ clients $c_i$ with $r_i = a_i$ for $1 \leq i \leq m$ and $r_{m+1} = 1$.
  \item $m+1$ internal nodes:\\
  - $m$ nodes $n_j$, $1 \leq j \leq m$, with $\W_j = \sto_j = a_j$.\\
  - A root node $r$ with $\W_r = \sto_r = S/2 + 1$.
  - The only child of $n_j$ is $c_j$. The parent of $n_j$ is $r$. The parent of $c_{n+1}$ is $r$.
\end{itemize}
Finally, we ask whether there exists a solution with total storage cost $S+1$.
Clearly, the size of $\II_2$ is polynomial (and even linear) in the size of $\II_1$.
We now show that instance $\II_1$ has a solution if and only if instance $\II_2$ does.
The same reduction works for both strategies, \CLOSEST and \MULTIPLE.

\begin{figure}
  \centering
  \includegraphics{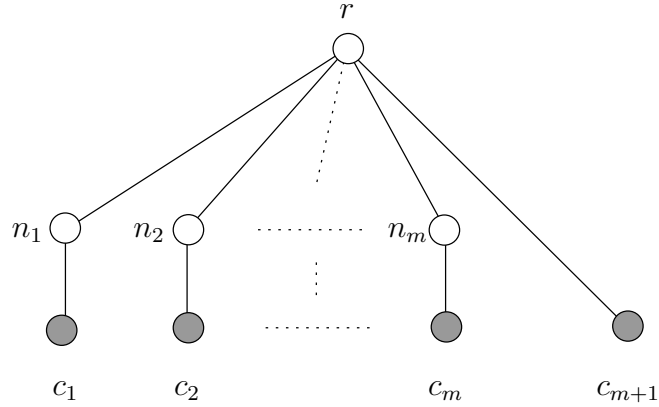}
  \caption{The platform used in the reduction for Theorem~\ref{th.nphetero}.}
  \label{fig.proof2}
\end{figure}

Suppose first that $\II_1$ has a solution. We assign a replica to each node $n_i$, $i \in I$, and
one in the root $r$. Client $c_i$ is served by $n_i$ if $i \in \II$, and by the root $r$ otherwise,
i.e. if $i \notin \II$ or if $i = m+1$. The total storage cost is $\sum_{j \in I} W_j + W_r = S +1$.
Because $\W_r = S/2 + 1 = \sum_{i \notin I} r_i + r_{n+1}$, the capacity of the root is not exceeded. Note that
the server allocation is compatible both with the \CLOSEST and \MULTIPLE
policies. In both cases, we have a solution to $\II_2$.

Suppose now that $\II_2$ has a solution. Necessarily, there is a replica located in the root,
otherwise client $c_{n+1}$ would not be served. Let $I$ be the index  set of nodes $n_j$,
$1 \leq j \leq n$, which have been allocated a replica in the solution of $\II_2$.
For $j \notin I$, there is no replica in node $n_j$, hence all requests of client $c_j$ are
processed by the root, whose storage capacity is $S/2 + 1$. We derive that $\sum_{j \notin I} r_j \leq S/2$.
Because the total storage capacity is $S+1$, the total storage capacity of nodes in $I$ is $S/2$.
The proof is slightly different for the two server strategies:
\begin{itemize}
  \item For the \CLOSEST strategy, all requests from a client $c_j \in I$
  are served by $n_j$, hence $\sum_{j \in I} r_j \leq S/2$.
  Since $\sum_{j \in I} r_j + \sum_{j \notin I} r_j =S$, we derive
  $\sum_{j \in I} r_j = \sum_{j \notin I} r_j =S/2$, hence a solution to $\II_2$.
  \item For the \MULTIPLE strategy, consider a server $j \in I$. Let $r'_j$ be the number of requests
  from client $c_j$ served by $n_j$, and $r''_j$ be the number of requests
  from $c_j$ served by the root $r$ (of course $r_j = r'_j + r''_j$).
  All requests from a
  client $c_j$, $j \notin I$, are served by the root.
  Let $A = \sum_{j \in I} r'_j$, $B = \sum_{j \in I} r''_j$ and $C = \sum_{j \notin I} r_j$.
  The total storage cost is $A+B+S/2+1$, hence $A+B \leq S/2$. We have seen that $C \leq S/2$.
  But $A+B+C=S$, hence $B=0$, and $A=C=S/2$, hence a solution to $\II_2$.
\end{itemize}

\end{proof}

\section{Linear programming formulation}
\label{sec.lp}

In this section, we express the \REP optimization problem in terms of an
integer linear program. We deal with the most general instance of the problem
on a heterogeneous tree, including QoS constraints,
and bounds on resource usage (both server and link
capacities). We derive a formulation for each of the three server
access policies, namely \CLOSEST, \UPWARDS and \MULTIPLE. This is an important extension
to a previous formulation due to~\cite{Karlsson02}.

While there is no efficient algorithm to solve integer linear
programs (unless P=NP), this formulation is extremely useful as it
leads to an absolute lower bound: we solve the integer linear
program over the rationals, using standard software
packages~\cite{Map,glpk}. Of course the rational solution will not
be feasible, as it assigns fractions of replicas to server nodes,
but it will provide a lower bound on the storage cost of any
solution. This bound will be very helpful to assess the performance
of the polynomial heuristics that are introduced in
Section~\ref{sec.heuristics}.

\subsection{Single server}

We start with single server strategies, namely
the \UPWARDS and \CLOSEST
access policies. We need to define a few variables:


\begin{description}
\item[Server assignment] $\;$
\begin{itemize}
  \item $x_j$ is a boolean variable equal to $1$ if $j$ is a server
(for one or several clients)
  \item $y_{i,j}$ is a boolean variable equal to $1$ if $j =
\server(i)$
  \item If $j \notin Ancests(i)$, we directly set $y_{i,j} = 0$.
\end{itemize}
\item[Link assignment] $\;$
\begin{itemize}
  \item $z_{i,l}$ is a boolean variable equal to $1$ if link $l \in
\pat{i}{r}$ is used when client $i$ accesses its server $\server(i)$
   \item If $l \notin \pat{i}{r}$ we directly set $z_{i,l} = 0$.
\end{itemize}
\end{description}

The objective function is the total storage cost, namely $\sum_{j \in \NN} \sto_j x_j$.
We list below the constraints common to the \CLOSEST and \UPWARDS policies:
First there are constraints for server and link usage:
\begin{itemize}
  \item Every client is assigned a server: $\forall i \in \CC, \sum_{j \in \Ancests(i)} y_{i,j} = 1$.

  \item All requests from $i\in \CC$ use the link to its parent: $z_{i, i \rightarrow \parent(i)} = 1$
  \item Let $i\in\CC$, and consider any link $l: j \rightarrow
j'=\parent(j) \in \pat{i}{r}$. If $j'=\server(i)$ then link $\suc(l)$
is not used by $i$ (if it exists). Otherwise $z_{i,\suc(l)}=z_{i,l}$. Thus:
  $$\forall i \in \CC, \forall l: j \rightarrow j'=\parent(j) \in \pat{i}{r},
  z_{i,\suc(l)} = z_{i,l} - y_{i,j'}$$

\end{itemize}

Next there are constraints expressing that server capacities and link bandwidths cannot be exceeded:
\begin{itemize}
  \item The processing capacity of any server cannot be exceeded:
  $\forall j \in \NN, \sum_{i \in \CC} r_i y_{i,j} \leq \W_j x_j.$
  Note that this ensures that if $j$ is the server of $i$, there is
indeed a replica located in node~$j$.
  \item The bandwidth of any link cannot be exceeded:
  $\forall l \in \LL, \sum_{i \in \CC} r_i z_{i,l} \leq \BW_l.$
\end{itemize}

Finally there remains to express the QoS constraints:
$$\forall i \in \CC, \forall j \in \Ancests(i), \dist(i,j) y_{i,j} \leq \qos_i,$$
where $\dist(i,j) = \sum_{l \in \pat{i}{j}} \comm_l.$ As stated
previously, we could take the computational time of a request into
account by writing $(\dist(i,j) + \comp_j) y_{i,j} \leq \qos_i$,
where $\comp_j$ would be the time to process a request on server
$j$.

Altogether, we have fully characterized the linear program for the \UPWARDS policy.
We need additional constraints for the \CLOSEST policy, which is a particular case of the
\UPWARDS policy (hence all constraints and equations remain valid).

We need to express that if node $j$ is the server of client $i$, then no ancestor of $j$ can be the server
of a client in the subtree rooted at $j$. Indeed, a client in this
subtree would need to be served by $j$
and not by one of its ancestors, according to the \CLOSEST policy. A direct
way to write this constraint is
$$\forall i \in \CC, \forall j \in \Ancests(i), \forall i' \in \CC \cap \subtree(j),
\forall j' \in \Ancests(j), y_{i,j} \leq 1 - y_{i',j'}.$$
Indeed, if $y_{i,j}=1$, meaning that $j= \server(i)$, then any client $i'$ in the subtree rooted in $j$
must have its server in that subtree, not closer to the root than $j$. Hence $y_{i',j'}=0$ for any
ancestor $j'$ of~$j$.

There are $O(s^4)$ such constraints to write, where $s = |\CC|+|\NN|$ is the problem size. We can reduce this
number down to $O(s^3)$ by writing
$$\forall i \in \CC, \forall j \in \Ancests(i)\setminus \{r\}, \forall i' \in \CC \cap \subtree(j),
y_{i,j} \leq 1 - z_{i',j \rightarrow \parent(j)}.$$

\subsection{Multiple servers}

We now proceed to the \MULTIPLE policy.
We define the following variables:

\begin{description}
\item[Server assignment] $\;$
\begin{itemize}
  \item $x_j$ is a boolean variable equal to $1$ if $j$ is a server
(for one or several clients)
  \item $y_{i,j}$ is an integer variable equal to the number of requests from client $i$
  processed by node~$j$
  \item If $j \notin Ancests(i)$, we directly set $y_{i,j} = 0$.
\end{itemize}
\item[Link assignment] $\;$
\begin{itemize}
  \item $z_{i,l}$ is an integer variable equal to the number of requests flowing through link $l \in
\pat{i}{r}$ when client $i$ accesses any of its servers in $\Servers(i)$
   \item If $l \notin \pat{i}{r}$ we directly set $z_{i,l} = 0$.
\end{itemize}
\end{description}

The objective function is unchanged, as the total storage cost still writes $\sum_{j \in \NN} \sto_j x_j$.
But the constraints must be modified. First those for server and link usage:
\begin{itemize}
  \item Every request is assigned a server: $\forall i \in \CC, \sum_{j \in \Ancests(i)} y_{i,j} = r_i$.
  \item All requests from $i\in\CC$ use the link to its parent: $z_{i, i \rightarrow \parent(i)} = r_i$
  \item Let $i\in\CC$, and consider any link $l: j \rightarrow
j'=\parent(j) \in \pat{i}{r}$. Some of the requests from $i$
  which flow through $l$ will be processed by node $j'$, and the
remaining ones will flow upwards through link $\suc(l)$:
  $$\forall i \in \CC, \forall l: j \rightarrow j'=\parent(j) \in \pat{i}{r},
  z_{i,\suc(l)} = z_{i,l} - y_{i,j'}$$
\end{itemize}

The other constraints on server capacities, link bandwidths and QoS are slightly modified:
\begin{itemize}
  \item Servers:
  $\forall j \in \NN, \sum_{i \in \CC}  y_{i,j} \leq \W_j x_j$. Note
that this ensure that if $j$ is the server for one or more requests
from $i$, there is indeed a replica located in node~$j$.
  \item Bandwidths: $\forall l \in \LL, \sum_{i \in \CC}  z_{i,l} \leq \BW_l$
  \item QoS: $\forall i \in \CC, \forall j \in \Ancests(i), \dist(i,j) y_{i,j} \leq \qos_i y_{i,j}$
\end{itemize}

Altogether, we have fully characterized the linear program for the \MULTIPLE policy.

\subsection{An ILP-based lower bound}
\label{sec.ilp}

The previous linear programs contain boolean or integer variables,
because it does not make sense to assign half a request or to place
one third of a replica on a node. However, we can still relax the
constraints and solve the linear program assuming that all variables
take rational values. The optimal solution of the relaxed program
can be obtained in polynomial time (in theory using the ellipsoid
method~\cite{Schrijver86}, in practice using standard software
packages~\cite{Map,glpk}), and the value of its objective function
provides an absolute lower bound on the cost of any valid (integer)
solution. Of course the relaxation makes the most sense for the
\MULTIPLE policy, because several fractions of servers are assigned
by the rational program. While not likely to be achievable, this
lower bound will provide an absolute reference for the performance
of the polynomial heuristics described in
Section~\ref{sec.heuristics}.

\section{Heuristics for the \COR problem}
\label{sec.heuristics}

In this section several heuristics for the \CLOSEST, \UPWARDS and
\MULTIPLE policies are presented. 
As previously stated, our main objective is to provide
an experimental assessment of the relative performance of the three
access policies. Our first attempt targets heterogenous trees without
QoS nor bandwidth constraints, thus considering the \COR problem, but further
work will be devoted to analyzing the impact of the additional
constraints (and in particular of the QoS constraints) on the
replica costs achieved by each policy.

All the eight heuristics described below have polynomial, and even
worst case quadratic complexity $O(s^2)$, where $s = |\CC|+|\NN|$ is
the problem size. Indeed, all heuristics proceed by traversing the
tree, 
and the number of traversals is bounded by the number of internal
nodes (and is much lower in practice).

We assume that each node $k \in \NN \cup \CC \setminus \{root\}$
knows its $\parent(k)$. Additionally, an internal node $j \in \NN$
knows its $\child(j)$, and the set $\clients(j)$ of the clients in
its subtree $\subtree(j)$. At any step of the heuristics, we denote
by $\inreq_j$ the number of requests in $\subtree(j)$ reaching $j$
with the current replicas already placed (initially, with no
replica, $\inreq_j = \sum_{i \in \clients(j)} r_i$). We use a
boolean variable $\treated_j$ to mark if a node $j$ has been treated
during a tree traversal. The set of replicas is initialized by
$\replica = \emptyset$.

\subsection{\CLOSEST}

The first two heuristics enforce the \CLOSEST policy through a
top-down approach, whereas the third heuristic uses a bottom-up
approach.

\paragraph{Closest Top Down All (\ctdall) --}
The basic idea is to perform a breadth-first traversal of the tree.
Every time a node is able to process the requests of all the clients
in its subtree, the node is chosen as a server, and we do not
explore further that subtree. The procedure
\emph{ClosestTopDownAll (\ctdall)} is presented in
Algorithm~\ref{algo:ctdall}. It is called until no more servers are
added in a tree traversal.

\begin{algorithm}[htbp]
  \caption{Procedure \ctdall}
  \label{algo:ctdall}
  \SetLine
  procedure {\bf \ctdall} (root, $\replica$)\\
    Fifo fifo\;
    fifo.push(root)\;
    \While{fifo $\neq \emptyset$}{
      $s$ = fifo.pop()\;
      \If{$s \notin \replica$}{
        \eIf{$W_s$ $\geq$ $\inreq_s$ $\&$ $\inreq_s$ $>$ 0}{
          $\replica$ = $\replica \cup \{s\}$\;

          \lForEach{$a \in \Ancests(s)$}{$\inreq_a = \inreq_a - \inreq_s$\;}
        }
        {
          \ForEach{$i \in \child(s)$}{
            \lIf{$i \in \NN$}{
              fifo.push($i$)\;
            }
          }
        }
      }
    }
\end{algorithm}

\paragraph{Closest Top Down Largest First (\ctdlf) --}
The tree is traversed in breadth-first manner as in \ctdall.
However, we treat the subtree which contains the most requests first
when considering the children of the tree (we sort the children by
increasing number of requests $\inreq$ to perform the
``fifo.push($i$)''). Also, instead of adding all possible servers in
a single step, the tree traversal is stopped as soon as a server
that can process all the requests in its subtree has been found.
This is done by adding an instruction $return$ each time a server
has been found in the procedure \ctdall
(Algorithm~\ref{algo:ctdall}), just after the update of the $\inreq$
values of the server's ancestors. As for the previous heuristic,
the procedure is called until no more server is chosen. In fact
\ctdlf is called exactly $|R|$ times, where $R$ is the final set of
replica.


\paragraph{Closest Bottom Up (CBU) --}
The last heuristic for the \CLOSEST policy performs a bottom-up
traversal of the tree. A node is chosen as a server if it can
process all the requests of the clients in its subtree.
Algorithm~\ref{algo:cbu} describes a recursive implementation of
\emph{ClosestBottomUp (\cbu)}. The procedure is initially called
with the root of the tree; while we do not reach the bottom of the
tree, we go down. Once arrived at the bottom, i.e. when the current
node~$s$ has only clients as children (test $atBottom(s)$) or when
all its children have already been treated (test
$allChildrenTreated(s)$), the node is marked as treated and added to
the set $\replica$ if $W_s \geq \inreq_s$. Then we go up in the tree
until all nodes are treated, performing recursive calls.

\begin{algorithm}[htbp]
  \caption{Procedure \cbu}
  \label{algo:cbu}
  \SetLine
  procedure {\bf \cbu} ($s \in \NN$, $\replica$)\\

      \eIf{atBottom($s$) $||$ allChildrenTreated($s$)}{
        $\treated_s$ = true\;
        \eIf{$W_s$ $\geq$ $\inreq_s$ $\&$ $\inreq_s$ $>$ 0}{
          /* node can treat all children's requests */\\
          $\replica = \replica \cup \{s\}$\;
          \lForEach{$a \in \Ancests(s)$}{$\inreq_a = \inreq_a - \inreq_s$\;}
        }
        {
          /* node cannot treat all children's requests, go up in the tree */\\
          \lIf{$\Ancests(s) \neq \emptyset$}{
            call \cbu (\parent($s$), $\replica$)\;
          }
        }

      }
        {
          \ForEach{$i \in \child(s)$}{
            /* not yet at the bottom of the tree, go down */\\
            \lIf{$i \in \NN\, \&$ $\neg \treated_i$}{
              call \cbu ($i$, $\replica$)\;
            }
          }
        }
\end{algorithm}

\paragraph{} Each of these three heuristics is placing a number of
replicas, but none is ensuring whether a valid solution has been
found or not. We need to check the final value of $\inreq_{root}$.
If there still are some pending requests at the root, there is no
valid solution. However, if $\inreq_{root}=0$, the heuristic has
found a solution.

\subsection{\UPWARDS}

We propose two heuristics for the \UPWARDS policy, the first one using
a top-down approach, the other considering the clients one by one, by
non-increasing order of their number of requests.

\paragraph{Upwards Top Down (UTD) --}

The top down approach works in two passes. In the first pass (see
Algorithm \ref{algo:utd1}), each node $s \in \NN$ whose capacity is
exhausted by the number of requests in its subtree ($W_s \leq
\inreq_s$) is chosen by traversing the tree in depth-first manner.
When a server is chosen, we delete as much clients as possible in
non-increasing order of their number of requests~$r_i$, until the
server capacity is reached or no other client can be deleted. This
delete procedure
is described in Algorithm~\ref{algo:delRequ}. 
If not all requests can be treated by the chosen servers, a second
pass is started. In this \emph{UTDSecondPass}-procedure (see
Algorithm \ref{algo:utd2}) servers with remaining requests are
added. Note that all these servers are non-exhausted by the
remaining requests ($\inreq_s < W_s$).
These two procedures are each called only once, with $s=root$ as a parameter.

Similarly to the \CLOSEST heuristics, we need to check that
$\inreq_{root}=0$ at the end of UTD to find out whether a valid
solution has been found.

\begin{algorithm}[htbp]
  \caption{Procedure deleteRequests}
  \label{algo:delRequ}
  \SetLine
  procedure {\bf deleteRequests} ($s \in \NN$, numToDelete)\\
  clientList = sortDecreasing(\clients($s$))\;
    \ForEach{$i \in$ clientList}{
      \If{$r_i$ $\leq$ numToDelete}{
        numToDelete = numToDelete - $r_i$\;
        \lForEach{$a \in \Ancests(i)$}{$\inreq_a = \inreq_a - r_i$\;}
        \child(\parent($i$)) = $\child(\parent(i)) \setminus \{i\}$\;
      \lIf{numToDelete == 0}{
        return\;
      }
    }
  }
\end{algorithm}

  \begin{algorithm}[htbp]
    \caption{Procedure \utdFirstPass}
    \label{algo:utd1}
    \SetLine
    procedure {\bf \utdFirstPass} ($s \in \NN$, $\replica$)\\

        \If{$\inreq_s$ $\geq$ $W_s$ $\&$ $\inreq_s$ $>$ 0}{
          $\replica = \replica \cup \{s\}$\;
          $\treated_s$ = true\;
          deleteRequests($s$, $W_s$)\;
        }
      \ForEach{$i \in \child(s)$}{
        \lIf{$i \in \NN$} { 
          \utdFirstPass($i$, $\replica$)\;
        }
      }

\end{algorithm}

  \begin{algorithm}[htbp]
    \caption{Procedure \utdSecondPass}
    \label{algo:utd2}
    \SetLine
    procedure {\bf \utdSecondPass} ($s \in \NN$, $\replica$)\\


      \eIf{$s \notin \replica\, \&$ $\inreq_s$ $>$ 0}{
        $\replica = \replica \cup \{s\}$\;
        deleteRequests($s$, $\inreq_s$)\;
      }
      {
       \ForEach{$i \in \child(s)$}{
          \lIf{$i \in \NN\, \& \, \inreq_i$ $>$ 0}{
            \utdSecondPass($i$, $\replica$)\;}
        }
      }
  \end{algorithm}


\paragraph{Upwards Big Client First (UBCF) --}
The second heuristic for the \UPWARDS policy works in a completely
different way than all the other heuristics. The basic idea here is
to treat all clients in non-increasing order of their $r_i$ values.
For each client we identify the server with minimal current capacity
(in the path from the client to the root) that can treat all its
requests. The capacity of a server is decreased each time it is
assigned some requests to process. If there is no valid server to
assign to a given client, the heuristic has failed to find a valid
solution. Please refer to Algorithm~\ref{algo:ubu} for details.

\begin{algorithm}[htbp]
  \caption{Procedure \ubu}
  \label{algo:ubu}
  \SetLine
  procedure {\bf \ubu} ($s \in \NN$, $\replica$)\\
  $clientList$ = sortDecreasing(\clients($s$)\;
  \ForEach{$i \in$ clientList}{
    $ValidAncests = \{ a \in \Ancests(i) | W_a \geq r_i \}$\;
    \If{$ValidAncests \neq \emptyset$}{
      $a = Min_{W_j} \{j \in ValidAncests\}$\;
    \lIf{$a \notin \replica$}{
          $\replica = \replica \cup \{a\}$\;
       }
       $W_a = W_a - r_i$\;
     }
     \lElse{return $no$ $solution$\;}
    }
 \end{algorithm}

\subsection{\MULTIPLE}

We propose three heuristics for the \MULTIPLE policy. The first one
uses a top-down approach, the second one a bottom-up approach. The
last one performs a greedy bottom-up traversal of the tree.

\paragraph{Multiple Top Down (\mtd) --}
The top-down approach for the \MULTIPLE policy is similar to the
top-down approach for \UPWARDS, with one significant difference: the
delete procedure. For \UPWARDS, requests of a client have to be
treated by a single server, and it may occur that after the
delete procedure a server still has some capacity left to treat more
requests, but all remaining clients have a higher amount of requests
than this leftover capacity. For \MULTIPLE, requests of a client can
be treated by multiple servers. So if at the end of the
delete procedure the server still has some capacity, we delete this
amount of requests from the client with the largest $r_i$.
This modified delete procedure is described in
Algorithm~\ref{algo:delRequMTD}.

\begin{algorithm}[htbp]
    \caption{Procedure deleteRequestsInMTD}
    \label{algo:delRequMTD}
  \SetLine
    procedure {\bf deleteRequestsInMTD} ($s \in \NN$, numToDelete)\\

    $clientList$ = sortDecreasing(\clients($s$))\;
      \ForEach{$i \in$ clientList}{
        \eIf{$r_i$ $\leq$ numToDelete}{
          numToDelete = numToDelete - $r_i$\;
          \lForEach{$a \in \Ancests(i)$}{$\inreq_a = \inreq_a - r_i$\;}
          \child(\parent($i$)) = $\child(\parent(i))\setminus \{i\}$\;
        }
        {
        $r_i$ =  $r_i$ - numToDelete\;
        \lForEach{$a \in \Ancests(i)$}{$\inreq_a = \inreq_a - r_i$\;}
        return\;
      }
    }
\end{algorithm}

\paragraph{Multiple Bottom Up (\mbu) --}
The first pass of this heuristic performs a bottom-up traversal of
the tree, as in \cbu. During this traversal,
nodes $s \in \NN$ are added to the set $\replica$ if their capacity
is exhausted ($W_s \leq \inreq_s$), similarly to the first pass of
the \mtd procedure. The delete procedure is identical to the \mtd
delete procedure (Algorithm~\ref{algo:delRequMTD}), except that
clients are deleted in non-decreasing order of their $r_i$ values
(instead of the non-increasing order). Intuitively, we aim at
deleting many small clients rather than fewer demanding ones.
The \emph{MBUFirstPass} is
described in Algorithm~\ref{algo:MBU1}, and the
\emph{MBUSecondPass}, which adds extra servers if required (similarly
to the second pass of \mtd), is
described in Algorithm~\ref{algo:MBU2}.

  \begin{algorithm}[htbp]
    \caption{Procedure \mbuFirstPass}
    \label{algo:MBU1}
      \SetLine
      procedure {\bf \mbuFirstPass} ($s \in \NN$, $\replica$)\\
          \eIf{atBottom($s$) $||$ allChildrenTreated($s$)}{
                $\treated_s$ = true\;
            \eIf{$W_s$ $\leq$ $inreq_s$ $\&$ $\inreq_s$ $>$ 0}{
              /* node is exhausted by the requests of its clients */\\
              $\replica = \replica \cup \{s\}$\;
              deleteRequestsInMBU($s$, $W_s$)\;
            }
            {
              /* node is not exhausted, go up the tree */\\
              \lIf{$\Ancests(s)\neq \emptyset$}{
                 call \mbu (\parent($s$), $\replica$)\;
              }
            }
          }
            {
              /* not yet at the bottom of the tree, go down */\\
              \ForEach{$i \in \child(s)$}{
                \lIf{$i \in \NN \,\&$ $\neg \treated_i$}{
                  call \mbu ($i$, $\replica$)\;
                }
             }
            }

\end{algorithm}

  \begin{algorithm}[htbp]
    \caption{Procedure \mbuSecondPass}
    \label{algo:MBU2}
    \SetLine
      procedure {\bf \mbuSecondPass} ($s \in \NN$, $\replica$)\\

        \eIf{$s \notin \replica$ $\&$ $\inreq_s$ $>$ 0}{
          $\replica = \replica \cup \{s\}$\;
          deleteRequestsInMBU($s$, $\inreq_s$)\;
        }
         { \ForEach{$i \in \child(s)$}{
            \lIf{$i \in \NN \, \& \, \inreq_i > 0 $}{
              \utdSecondPass($i$, $\replica$)\;
            }
          }
        }


\end{algorithm}

%

\paragraph{Multiple Greedy (\mg) --}
The last heuristic performs a greedy bottom-up assignment of
requests, similarly to
Pass~3 of the optimal algorithm for the homogeneous case (see
Algorithm~\ref{alg:pass3} in Section~\ref{sec.elegant}). We add a
replica whenever there are some requests affected to a server. For
heterogeneous platforms, we may often return a cost far from the
optimal, but we ensure that we always find a solution to the problem
if there exists one.

It might be particularly interesting to use \mg only for problem
instances for which \mbu or \mtd fail to find a solution.

\section{Experiments: comparisons of different access policies}
\label{sec.experiments}

We have done some experiments to assess the impact of the different
access policies, and the performance of the polynomial heuristics
described in Section~\ref{sec.heuristics}. We obtain an absolute
lower bound of the solution for each tree platform with a linear
program similar to those of Section~\ref{sec.lp}, but modified so as
to solve larger problems. Section~\ref{exp.lp} details how we
compute this lower bound. We outline the experimental plan in
Section~\ref{exp.plan}. Results are given and commented in
Section~\ref{exp.res}. In the following, we denote by $s$ the
problem size: $s = |\CC|+|\NN|$.

\subsection{Obtaining a lower bound}
\label{exp.lp}

The linear programs exposed in Section~\ref{sec.lp} must be solved
in integer values if we wish to obtain an exact solution to an
instance of the problem. This can be done for each access policy,
but due to the large number of variables, the problem cannot be
solved for platforms of size $s>50$. Thus we cannot use this
approach for large-scale problems.

For all practical values of the problem size, the rational linear
program returns a solution in a few minutes. We tested up to several
thousands of nodes and clients, and we always found a solution within
ten seconds.

However, we can obtain a more precise lower bound for trees with up
to $s=400$ nodes and clients by using a rational solution of the
\MULTIPLE instance of the linear program with fewer integer
variables. We treat the $y_{i,j}$ and $z_{i,l}$ as rational
variables, and only require the $x_j$ to be integer variables. These
variables are set to $1$ if and only if there is a replica on the
corresponding node. Thus, forbidding to set $0<x_j<1$ allows us to
get a realistic value of the cost of a solution of the problem. For
instance, a server might be used only at 50\% of its capacity, thus
setting $x=0.5$ would be enough to ensure that all requests are
processed; but in this case, the cost of placing the replica at this
node is halved, which is incorrect: while we can place a replica or
not but it is impossible to place half of a replica.

In practice, this lower bound provides a drastic improvement over
the unreachable lower bound provided by the fully rational linear
program. The good news is that we can compute the refined lower
bound for problem sizes up to $s=400$, using GLPK~\cite{glpk}. We
used the refined bound for all our experiments.

\subsection{Experimental plan}
\label{exp.plan}

The important parameter in our tree networks is the load, i.e. the
total number of requests compared to the total processing power:
$$\lambda = \frac{\sum_{i \in \CC} \rr_i}{\sum_{j \in \NN} \W_i}$$

We have performed experiments on $30$ trees for each of the nine
values of $\lambda$ selected ($\lambda = 0.1, 0.2, ..., 0.9$). The
trees have been randomly generated, with a problem size $15 \leq s
\leq 400$. When $\lambda$ is small, the tree has a light request
load, while large values of $\lambda$ implies a heavy load on the
servers. We then expect the problem to have a solution less
frequently.

We have computed the number of solutions for each lambda and each
heuristic. The number of solutions obtained by the linear program
indicates which problems are solvable. Of course we cannot expect a
result with our heuristics for those intractable problems.

To assess the relative cost of each heuristic, we have studied the
distance of the result (in terms of replica cost) of the heuristic
to the lower bound. This allows to compare the cost of the different
heuristics, and thus to compare the different access policies. For
each $\lambda$, the cost is computed on the trees for which the
linear program has a solution. Let $T_\lambda$ be the subset of
trees with a solution. Then, the relative cost for the heuristic $h$
is obtained by:
$$rcost = \frac{1}{|T_\lambda|} \sum_{t \in T_\lambda}
\frac{cost_{LP}(t)}{cost_h(t)}$$
where $cost_{LP(t)}$ is the lower bound cost returned by the linear
program on tree $t$, and $cost_h(t)$ is the cost involved by the
solution proposed by heuristic~$h$. In order to be fair versus
heuristics who have a higher success rate, we set $cost_h(t)=+\infty$
if the heuristic did not find any solution.

Experiments have been conducted both on homogeneous networks (\RCO
problem) and on heterogeneous ones (\COR problem).

\subsection{Results}
\label{exp.res}



A solution computed by a \CLOSEST or \UPWARDS heuristic always is a
solution for the \MULTIPLE policy, since the latter is less
constrained. Therefore, we can mix results into a new heuristic for
the \MULTIPLE policy, called MixedBest ({\bf MB}), which selects for
each tree the best cost returned by the previous eight heuristics
for this particular problem instance. Since MG never fails to find a
solution if there is one, MB will neither fail either.

\begin{figure}[phbt]
\begin{center}
\includegraphics[angle=270,width=12cm]{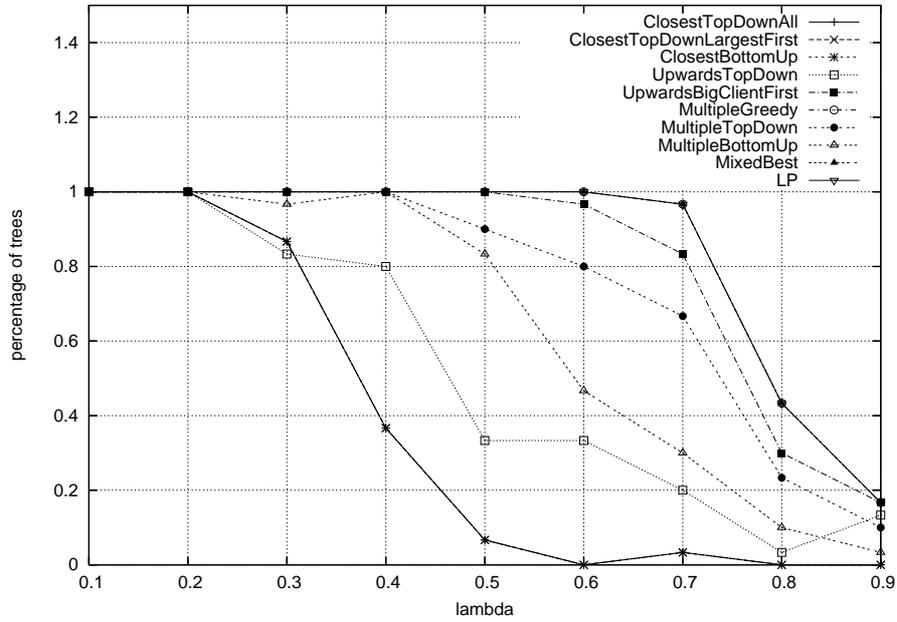}
\end{center}
\caption{Homogeneous case - Percentage of success.}
\label{fig.perc-hom}
\end{figure}

Figure~\ref{fig.perc-hom} shows the percentage of success of each
heuristic for homogeneous platforms. The upper curve corresponds to
the result of the linear program, and to the cost of the MG and MB
heuristics, which confirms that they always find a solution when
there is one. The \ubu heuristic seems very efficient, since it
finds a solution more often than MTD and MBU, the other two
\MULTIPLE policies. On the contrary, \utd, which works in a similar
way to \mtd and \mbu, finds less solutions than these two
heuristics, since it is further constrained by the \UPWARDS policy.
As expected, all the \CLOSEST heuristics find fewer solutions as
soon as $\lambda$ reaches higher values: the bottom curve of the
plot corresponds to \ctdall, \ctdlf and \cbu, which all find the
same solutions. This is inherent to the limitation of the \CLOSEST
policy: when the number of requests is high compared to the total
processing power in the tree, there is little chance that a server
can process all the requests coming from its subtree, and requests
cannot traverse this server to be served higher in the tree. These
results confirm that the new policies have a striking impact on the
existence of a solution to the \RCO problem.

\begin{figure}[phbt]
\begin{center}
\includegraphics[angle=270,width=12cm]{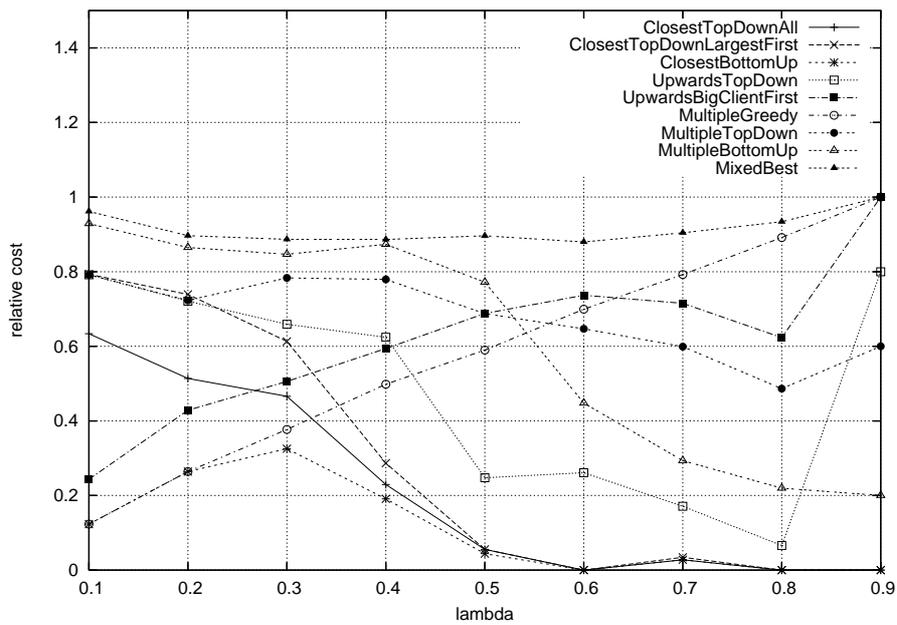}
\end{center}
\caption{Homogeneous case - Relative cost.}
\label{fig.cost-hom}
\end{figure}

Figure~\ref{fig.cost-hom} represents the relative cost of the
heuristics compared to the LP-based lower bound.
As expected, the hierarchy between the policies is respected, i.e.
\MULTIPLE is better than \UPWARDS which in turn is better than
\CLOSEST. For small values of $\lambda$, it happens that some
\CLOSEST heuristics give a better solution than those for \UPWARDS
or \MULTIPLE, due to the fact that the latter heuristics are not
well optimized for small values of $\lambda$. Also, \ubu is better
than all the \MULTIPLE heuristics for $\lambda=0.6$. Altogether, the
use of the MixedBest heuristic MB allows to always pick up the best
result, thereby resulting in a very satisfying relative cost for the
\MULTIPLE instance of the problem. The greedy MG should not be used
for small values of $\lambda$, but proves to be very efficient for
large values, since it is the only heuristic to find a solution for
such instances.
%
To conclude, we point out that MB always achieves a relative cost of
at least 85\%, thus returning a replica cost within $17\%$ of that
of the LP-based lower bound. This is a very satisfactory result for
the absolute performance of our heuristics.

\begin{figure}[phbt]
\begin{center}
\includegraphics[angle=270,width=12cm]{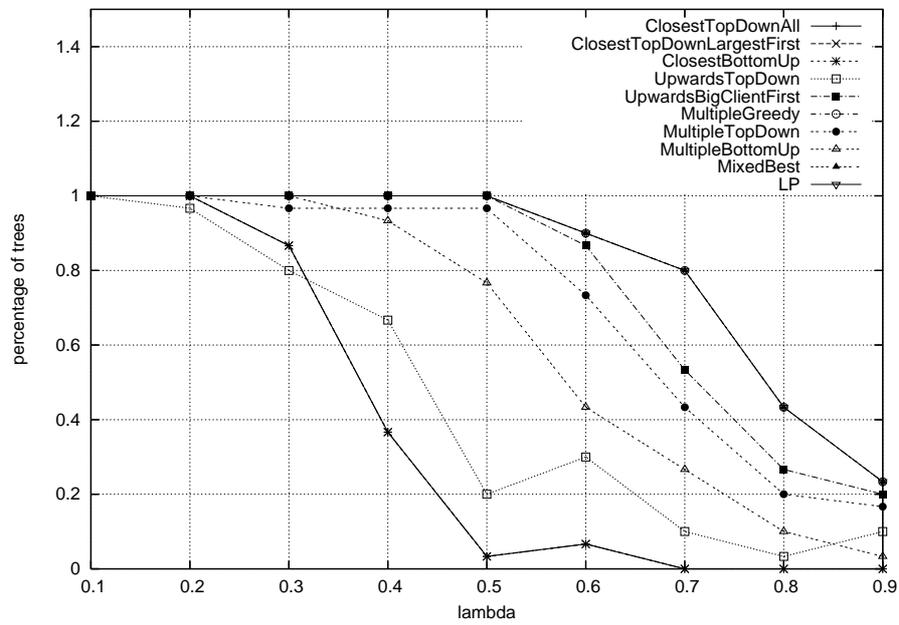}
\end{center}
\caption{Heterogeneous case - Percentage of success.}
\label{fig.perc-het}
\end{figure}

\begin{figure}[phbt]
\begin{center}
\includegraphics[angle=270,width=12cm]{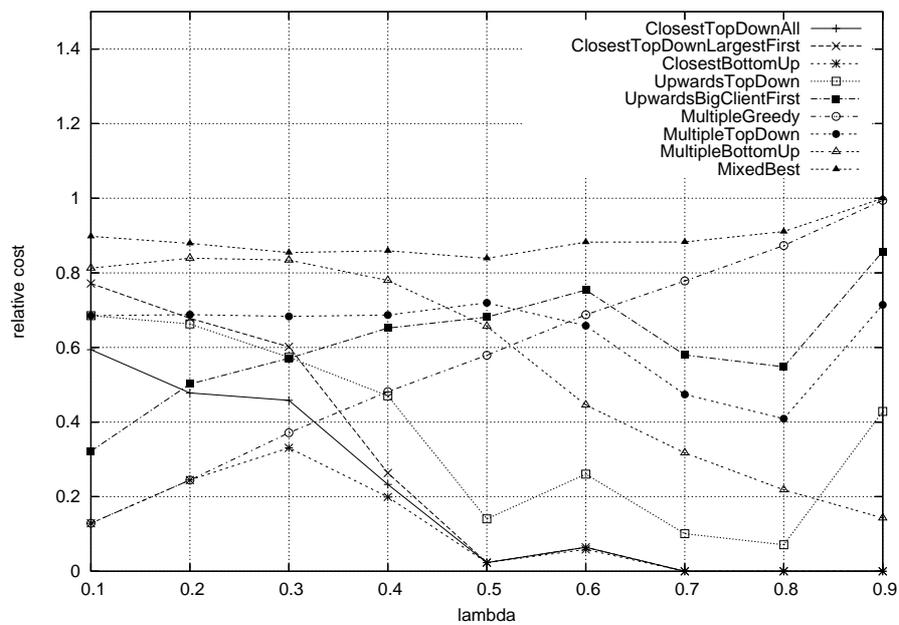}
\end{center}
\caption{Heterogeneous case - Relative cost.} \label{fig.cost-het}
\end{figure}

The heterogeneous results (see Figure~\ref{fig.perc-het} and
Figure~\ref{fig.cost-het})  are very similar to the homogeneous
ones, which clearly shows that our heuristics are not much sensitive
to the heterogeneity of the platform. Therefore, we have an
efficient way to find in polynomial time a good solution to all the
NP-hard problems stated in Section~\ref{sec.complexity}.

\section{Extensions}
\label{sec.extensions}

In this paper we have considered a simplified instance of the
replica problem.
In this section, we outline two important generalizations, namely
dealing with several objects, and changing the
objective function.

%

\subsection{With several objects}
\label{sec.wso}

In this paper, we have restricted the study of the problem to a
single object, which means that all replicas are identical (of the same type). We
can envision a system in which different types of objects need to be
accessed. The clients are then having requests of different types, which
can be served only by an appropriate replica. Thus, for an object of
type~$k$, client $i\in \CC$ issues $\rr_i^{(k)}$ requests for this
object. To serve a request of type~$k$, a node must be provided
with a replica of that type. Nodes can be provided with several replica types.
A given client is likely to have different servers for different objects.
The QoS may also be object-dependent ($\qos_i^{(k)}$).

To refine further, new parameters
can be introduce such as the size of object~$k$ and the
computation time involved for this object. Nodes parameters become
object-dependent too, in particular the storage cost and the time
required to answer a request.

The server capacity constraint must then be a sum on all the object
types, while the QoS must be satisfied for each object type. The link
capacity also is a sum on the different object types, taking into
account the size of each object.

There remains to modify the objective function: we simply aim at minimizing
the cost of all replicas of different types that have been assigned
to the nodes in the solution to get
the extended \emph{replica cost} for several objects.

Because the constraints add up linearly for different objects, it is not difficult
to extend the linear programming formulation of Section~\ref{sec.lp}
to deal with several objects. Also, the three access policies \CLOSEST,
\UPWARDS and \MULTIPLE could naturally be extended to handle several objects.
However, designing efficient
heuristics for various object types, especially with different communication to computation
ratios and different QoS constraints for each type, is a challenging algorithmic problem.

\subsection{More complex objective functions}
\label{sec.mcof}

Several important extensions of the problem
consist in having a more complex objective function. In fact,
either with on or with several objects,
we have restricted so far to minimizing the cost of the replicas (and even their
number in the homogeneous case). However,
several other factors can be introduced in the objective function:

\begin{description}
  \item[Communication cost --] This cost is the {\em read} cost, {\em
i.e.} the communication cost required to access the replicas to answer
requests. It is thus a sum on all objects and all clients of the
communication time required to access the replica. If we take this
criteria into account in the objective function, we may prefer a
solution in which replicas are close to the clients.

  \item[Update cost --] The {\em write} cost is the extra cost
due to an update of the replicas. An update must be performed when one
of the clients is modifying (writing) some of the data. In this case,
to ensure the consistency of the data, we need to propagate the
modification to all other replicas of the modified object. Usually,
this cost is directly related to the communication costs on the
minimum spanning tree of the replica, since the replica which has been
modified sends the information to all the other replicas.

  \item[Linear combination --] A quite general objective function
can be obtained by a linear combination of the three different costs,
namely replica cost, read cost and write cost. Informally, such an objective function
would write
$$ \alpha \sum_{\text{servers, objects}} \textit{replica cost} + \;
\beta \sum_{\text{requests}} \textit{read cost} + \;
\gamma
\sum_{\text{updates}} \textit{write cost}$$ where the
application-dependent parameters $\alpha$, $\beta$ and $\gamma$
would be used to give priorities to the different costs.
  \end{description}

Again, designing efficient
heuristics for such general objective functions, especially in the context
of heterogeneous resources, is a challenging algorithmic problem.

\section{Related work}
\label{sec.related}

Early work on replica placement by Wolfson and Milo~\cite{Wolfson91}
has shown the impact of the write cost and motivated the use of a
minimum spanning tree to perform updates between the replicas. In
this work, they prove that the replica placement problem in a
general graph is NP-complete, even  without taking into account
storage costs. Thus they address the case of special topologies, and
in particular tree networks. They give a polynomial solution in a
fully homogeneous case and a simple model with no QoS and no server
capacity. Their work uses the closest server access policy (single
server) to access the data. 

Using this \CLOSEST policy, Cidon et al~\cite{Cidon2002} studied an
instance of the problem with
multiple objects. In this work, the objective function
has no update cost, but integrates a communication cost.
Communication cost in the objective function can be seen as a
substitute for QoS. Thus, they minimize the average
communication cost for all the clients rather than ensuring a given
QoS for each client. They target fully homogeneous platforms since
there are no server capacity constraints in their approach. A
similar instance of the problem has been studied by Liu et
al~\cite{PangfengLiu06}, adding a QoS in terms of a range limit
(QoS=distance), and the objective being the \RCO problem. In this
latter approach, the servers are homogeneous, and their capacity is
bounded.

Cidon et al~\cite{Cidon2002} and Liu et al~\cite{PangfengLiu06} both
use the \CLOSEST access policy. In each case, the optimization
problems are shown to have polynomial complexity. However, the
variant with bidirectional links is shown NP-complete by Kalpakis et
al~\cite{Kalpakis2001}. Indeed in~\cite{Kalpakis2001}, requests can
be served by any node in the tree, not just the nodes located in the
path from the client to the root. The simple problem of minimizing
the number of replicas with identical servers of fixed capacity,
without any communication cost nor QoS contraints, directly reduces
to the clasical bin packing problem.

Kalpakis et al~\cite{Kalpakis2001} show that a special instance of
the problem is polynomial, when considering no server capacities,
but with a general objective function taking into account read,
write and storage costs. In their work, a minimum spanning tree is
used to
propagate the writes, as was done in~\cite{Wolfson91}. 
Different methods can however be used, such as a minimum cost
Steiner tree, in order to further optimize the write
strategy~\cite{Kalpakis-st01}.

All papers listed above consider the \CLOSEST access policy. As
already stated, most problems are NP-complete, except for some very
simplified instances. Karlsson et al~\cite{Karlsson02,Karlsson04}
compare different objective functions and several heuristics to
solve these complex problems. They do not take QoS constraints into
account, but instead integrate a communication cost in the objective
function as was done in~\cite{Cidon2002}. Integrating the
communication cost into the objective function can be viewed as a
Lagrangian relaxation of QoS constraints.

Tang and Xu~\cite{Tang2005} have been one of the first authors to
introduce actual QoS constraints in the problem formalization. In
their approach, the QoS corresponds to the latency requirements of
each client. Different access policies are considered. First, a
replica-aware policy in a general graph is proven to be NP-complete.
When the clients do not know where the replicas are (replica-blind
policy), the graph is simplified to a tree (fixed routing scheme)
with the \CLOSEST policy, and in this case again it is possible to
find a polynomial algorithm using dynamic programming.

\medskip
To the best of our knowledge, there is no related work comparing
different access policies, either on tree networks or on general
graphs. Most previous works impose the \CLOSEST policy.
The \MULTIPLE policy is enforced by Rodolakis et
al~\cite{Rodolakis06} but in a very different context. In fact, they
consider general graphs instead of trees, so they face the
combinatorial complexity of finding good routing paths. Also, they
assume an unlimited capacity at each node, since they can add
numerous servers of different kinds on a single node. Finally, they
include some QoS constraints  in their problem formulation, based on
the round trip time (in the graph) required to serve the client
requests. In such a context, this (very particular) instance of the
\MULTIPLE problem is shown to be NP-hard.

\section{Conclusion}
\label{sec.conclusion}

In this paper, we have introduced and extensively analyzed two
important new policies for the replica placement problem. The
\UPWARDS and \MULTIPLE policies are natural variants of the standard
\CLOSEST approach, and it may seem surprising that they have not
already been considered in the published literature.

On the theoretical side, we have fully assessed the complexity of
the \CLOSEST, \UPWARDS and \MULTIPLE policies, both for homogeneous
and heterogeneous platforms. The polynomial complexity of the
\MULTIPLE policy in the homogeneous case is quite unexpected, and we
have provided an elegant algorithm to compute the optimal cost for
this policy. Not surprisingly, all three policies turn out to be
NP-complete for heterogeneous nodes, which provides yet another
example of the additional difficulties induced by resource
heterogeneity.

On the practical side, we have designed several heuristics for the
\CLOSEST, \UPWARDS and \MULTIPLE policies, and we have compared
their performance for a simple instance of the problem, without QoS
constraints nor bandwidth limitations. In the experiments, the
constraints were only related to server capacities, and the total
cost was the sum of the server capacities (or their number in the
homogeneous case). Even in this simple setting, the impact of the
new policies is impressive:
the number of trees which admit a solution is much higher with the
\UPWARDS and \MULTIPLE policies than with the \CLOSEST policy.
Finally,
we point out that the absolute performance of the heuristics is
quite good, since their cost is close to the lower bound based upon
the solution of the integer linear program.

\medskip
There remains much work to extend the results of this paper, in
several important directions. In the short term, we need to conduct
more simulations for the \COR problem, varying the shape of the
trees, the distribution law of the requests and the degree of
heterogeneity of the platforms. We also aim at designing efficient
heuristics for more general instances of the \REP problem, taking
QoS and bandwidth constraints into account. It will be instructive
to see whether the superiority of the new \UPWARDS and \MULTIPLE
policies over \CLOSEST remains so important in the presence of QoS
constraints. Also, including bandwidth constraints may require a
better global load-balancing along the tree, thereby favoring
\MULTIPLE over \UPWARDS.

In the longer term, designing efficient heuristics for the problem
with various object types, all with different communication to
computation ratios and different QoS constraints is a demanding
algorithmic problem. Also, we would like to extend this work so as
to handle more complex objective functions, including communication
costs and update costs as well as replica costs; this seems to be a
very difficult challenge to tackle, especially in the context of
heterogeneous resources.

\bibliographystyle{abbrv}
\bibliography{biblio}

\begin{thebibliography}{10}

\bibitem{Map}
B.~W. Char, K.~O. Geddes, G.~H. Gonnet, M.~B. Monagan, and S.~M. Watt.
\newblock {\em Maple Reference Manual}, 1988.

\bibitem{Cidon2002}
I.~Cidon, S.~Kutten, and R.~Soffer.
\newblock Optimal allocation of electronic content.
\newblock {\em Computer Networks}, 40:205--218, 2002.

\bibitem{GareyJohnson}
M.~R. Garey and D.~S. Johnson.
\newblock {\em Computers and Intractability, a Guide to the Theory of
  {NP}-Completeness}.
\newblock W.H. Freeman and Company, 1979.

\bibitem{glpk}
{GLPK: GNU Linear Programming Kit}.
\newblock \url{http://www.gnu.org/software/glpk/}.

\bibitem{Kalpakis2001}
K.~Kalpakis, K.~Dasgupta, and O.~Wolfson.
\newblock Optimal placement of replicas in trees with read, write, and storage
  costs.
\newblock {\em IEEE Trans. Parallel and Distributed Systems}, 12(6):628--637,
  2001.

\bibitem{Kalpakis-st01}
K.~Kalpakis, K.~Dasgupta, and O.~Wolfson.
\newblock {Steiner-Optimal Data Replication in Tree Networks with Storage
  Costs}.
\newblock In {\em {IDEAS '01}: Proceedings of the 2001 International Symposium
  on Database Engineering \& Applications}, pages 285--293. IEEE Computer
  Society Press, 2001.

\bibitem{Karlsson04}
M.~Karlsson and C.~Karamanolis.
\newblock {Choosing Replica Placement Heuristics for Wide-Area Systems}.
\newblock In {\em ICDCS '04: Proceedings of the 24th International Conference
  on Distributed Computing Systems (ICDCS'04)}, pages 350--359, Washington, DC,
  USA, 2004. IEEE Computer Society.

\bibitem{Karlsson02}
M.~Karlsson, C.~Karamanolis, and M.~Mahalingam.
\newblock A framework for evaluating replica placement algorithms.
\newblock Research Report HPL-2002-219, HP Laboratories, Palo Alto, CA, 2002.

\bibitem{PangfengLiu06}
P.~Liu, Y.-F. Lin, and J.-J. Wu.
\newblock Optimal placement of replicas in data grid environments with locality
  assurance.
\newblock In {\em International Conference on Parallel and Distributed Systems
  (ICPADS)}. IEEE Computer Society Press, 2006.

\bibitem{Rodolakis06}
G.~Rodolakis, S.~Siachalou, and L.~Georgiadis.
\newblock {Replicated server placement with {QoS} constraints}.
\newblock {\em IEEE Trans. Parallel Distributed Systems}, 17(10):1151--1162,
  2006.

\bibitem{Schrijver86}
A.~Schrijver.
\newblock {\em Theory of Linear and Integer Programming}.
\newblock John Wiley \& Sons, New York, 1986.

\bibitem{Tang2005}
X.~Tang and J.~Xu.
\newblock {QoS-Aware Replica Placement for Content Distribution}.
\newblock {\em IEEE Trans. Parallel Distributed Systems}, 16(10):921--932,
  2005.

\bibitem{Wolfson91}
O.~Wolfson and A.~Milo.
\newblock The multicast policy and its relationship to replicated data
  placement.
\newblock {\em ACM Trans. Database Syst.}, 16(1):181--205, 1991.

\end{thebibliography}

\end{document}